\documentclass[11pt]{article}
\usepackage{amsfonts}
\usepackage{amssymb}
\usepackage{amsmath}
\usepackage{a4wide}
\usepackage{graphicx}

\newtheorem{theorem}{Theorem}

\newtheorem{corollary}{Corollary}

\newtheorem{lemma}{Lemma}

\newenvironment{proof}[1][Proof]{\textbf{#1.} }{\ \rule{0.5em}{0.5em}}
\makeatletter
\def\@removefromreset#1#2{\let\@tempb\@elt
     \def\@tempa#1{@&#1}\expandafter\let\csname @*#1*\endcsname\@tempa
     \def\@elt##1{\expandafter\ifx\csname @*##1*\endcsname\@tempa\else
    \noexpand\@elt{##1}\fi}     \expandafter\edef\csname cl@#2\endcsname{\csname cl@#2\endcsname}     \let\@elt\@tempb
     \expandafter\let\csname @*#1*\endcsname\@undefined}

\@removefromreset{equation}{section}

\@removefromreset{theorem}{section}
\makeatother
\input{tcilatex}

\begin{document}

\title{Multipartite Bell-type inequalities for arbitrary numbers of settings
and outcomes per site}
\author{Elena R. Loubenets\thanks{%
E-mail: erl@erl.msk.ru}\bigskip \\
Applied Mathematics Department, Moscow State Institute \\
of Electronics and Mathematics, Moscow 109028, Russia}
\maketitle

\begin{abstract}
We introduce a single general representation incorporating in a unique
manner all Bell-type inequalities for a multipartite correlation scenario
with an arbitrary number of settings and any spectral type of outcomes at
each site. Specifying this general representation for correlation functions,
we prove that the form of any correlation Bell-type inequality does not
depend on a spectral type of outcomes, in particular, on their numbers at
different sites, and is determined only by extremal values of outcomes at
each site. We also specify the general form of bounds in Bell-type
inequalities on joint probabilities. Our approach to the derivation of
Bell-type inequalities is universal, concise and can be applied to a
multipartite correlation experiment with outcomes of any spectral type,
discrete or continuous. We, in particular, prove that, for an N-partite
quantum state, possibly, infinite dimensional, admitting the $\underbrace{%
2\times ...\times 2}_{N}$-setting LHV description, the Mermin-Klyshko
inequality holds for any two bounded quantum observables per site, not
necessarily dichotomic.
\end{abstract}

\tableofcontents

\section{ Introduction}

A Bell-type inequality represents a tight\footnote{%
In the present paper, the term \textit{a tight }LHV constraint means that,
in the LHV frame, the bounds established by this constraint cannot be
improved. On the difference between the terms \textit{a tight linear LHV} 
\textit{constraint} and \textit{an extreme} \textit{linear LHV constraint},
see the end of section 2.1.} linear probabilistic constraint on correlation
functions or joint probabilities that holds under any multipartite
correlation experiment admitting a local hidden variable (LHV) description
and may be violated otherwise. Proposed first [1-3] as tests on the
probabilistic description of quantum measurements, these inequalities are
now widely used in many quantum information schemes and have been
intensively discussed in the literature.

Nevertheless, the most analysed versions [4-18] of Bell-type inequalities
refer to either a multipartite case with two settings and two outcomes per
site or a bipartite case with small numbers of settings and outcomes and we
still know a little about Bell-type inequalities for an arbitrary
multipartite correlation experiment. Note, however, that a generalized
quantum measurement on even a qubit may have infinitely many outcomes.

In the literature on quantum information, finding of Bell-type inequalities
for larger numbers of settings and outcomes per site is considered to be a
computationally hard problem. This is really the case in the frame of the
generally accepted polytope approach [19] where the construction of a
complete set of extreme Bell-type inequalities is associated with finding
of\ \emph{all} faces of a highly dimensional polytope. However, many of
these faces correspond to \emph{trivial}\footnote{%
In the sense that these constraints hold under any multipartite correlation
experiment, not necessarily admitting an LHV description.} probabilistic
constraints while others can be subdivided into only a few classes, each
describing extreme Bell-type inequalities of the same form. It was also
shown [17] computationally that increasing of numbers of settings and
outcomes per site, resulting in the appearance of a huge amount of new
faces, leads to only a few (or possibly, no any) new forms of extreme
Bell-type inequalities for joint probabilities. Moreover, in case of an
inifinite number of outcomes per site, the polytope approach cannot be, in
principle, used for the construction of Bell-type inequalities on joint
probabilities of arbitrary events, not necessarily of the product form.

The problem is also complicated by the fact that Bell-type inequalities for
correlation functions and Bell-type inequalities for joint probabilities are
usually considered separately and a \emph{general} link between the forms of
these inequalities in an arbitrary multipartite case has not been analysed
in the literature\footnote{%
In a $2\times 2$-setting case, this link was considered by Fine [4] for two
outcomes per site and by Masanes [15] for $d\geq 2$ discrete outcomes at
each site.}.

In the present paper, which is a sequel to\footnote{%
In [20], we have consistently formalized the probabilistic description of a
multipartite correlation experiment, performed on systems of any nature and
with outcomes of any spectral type, discrete or continuous.} [20], we make a
step in this direction by introducing a \emph{single }general representation
(theorem 1, section 2), incorporating in a unique manner \emph{all} tight
linear LHV constraints on either correlation functions or joint
probabilities arising under an $S_{1}\times ...\times S_{N}$-setting $N$%
-partite correlation experiment with outcomes of \emph{any} spectral type,
discrete or continuous.

Specifying\ this general representation for correlation functions, we prove
(corollaries 1, 2, section 2.1) that the form of any \emph{correlation}
Bell-type inequality \emph{does not depend} on a spectral type of outcomes
observed at different sites and is determined only by extremal values of
outcomes at each site.

The general form of bounds in tight linear LHV constraints on joint
probabilities is specified by corollaries 3, 4 in section 2.2.

All Bell-type inequalities that have been introduced in the literature
[1-18] constitute particular cases of this single general representation. We
explicitly demonstrate (section 3) this for: (a) the
Clauser-Horne-Shimony-Holt (CHSH) inequality [2] for correlation functions;
(b) the Clauser-Horne (CH) inequalities [3] for joint probabilities; (c) the
Mermin-Klyshko (MK) inequality [6-8] for correlation functions; (d) the
Bell-type inequalities for joint probabilities found computationally by
Collins and Gisin [17]; (e) the Zohren-Gill inequality [18] for joint
probabilities.

Our approach to the derivation of Bell-type inequalities is universal,
concise and allows us to extend the applicability ranges of even the
well-known Bell-type inequalities. Applying, for example, this appoach to an 
$N$-partite correlation experiment, with two settings and any spectral type
of outcomes at each site, we derive the Bell-type inequality (section 3.3)
that, being specified for a quantum case, takes the form of the
Mermin-Klyshko (MK) inequality [6-8] for spin measurements on $N$ qubits.
This proves that, for a quantum state $\rho $ on $\mathcal{H}_{1}\otimes
...\otimes \mathcal{H}_{N},$ admitting the $\underbrace{2\times ...\times 2}%
_{N}$-setting LHV description, the MK inequality holds for any two bounded
quantum observables per site, not necessarily dichotomic. If a Hilbert space 
$\mathcal{H}_{n}$, corresponding to $n$-th site, is infinite dimensional
then bounded quantum observables measured at this site may be of any
spectral type, discrete or continuous.

\section{Linear LHV constraints}

Consider an $N$-partite correlation experiment where an $n$-th party
performs $S_{n}\geq 1$ measurements, each specified by a positive integer $%
s_{n}\in \{1,...,S_{n}\}\ $and with outcomes $\lambda _{n}^{(s_{n})}\in
\Lambda _{n}^{(s_{n})}$ of any spectral type, discrete or continuous, not
necessarily real numbers.

This correlation experiment is described by the $S_{1}\times ...\times S_{N}$%
-setting family\footnote{%
On details of notation, see sections 2, 3 of [20].}%
\begin{equation}
\mathcal{E}=\{(s_{1},...,s_{N})\mid s_{1}=1,...,S_{1},...,s_{N}=1,...,S_{N}\}
\label{1}
\end{equation}%
of $N$-partite joint measurements with joint probability distributions%
\begin{equation}
P_{(s_{1},...,s_{_{N}})}^{(\mathcal{E)}}(\mathrm{d}\lambda
_{1}^{(s_{1})}\times ...\times \mathrm{d}\lambda _{N}^{(s_{_{N}})}),\text{ \
\ \ }s_{1}=1,...,S_{1},...,s_{N}=1,...,S_{N},  \label{2}
\end{equation}%
where each distribution $P_{(s_{1},...,s_{_{N}})}^{(\mathcal{E)}}$ may, in
general, depend not only on settings of the corresponding joint measurement $%
(s_{1},...,s_{_{N}})$ but also on a structure of the whole experiment $%
\mathcal{E}.$

For an $N$-partite joint measurement $(s_{_{1}},...,s_{_{N}})\in \mathcal{E}%
, $ let us denote by\footnote{%
For an integral taken over all values of variables, the domain of
integration is not usually specified.}%
\begin{equation}
\langle \text{ }\Psi (\lambda _{1}^{(s_{_{1}})},...,\lambda
_{N}^{(s_{_{N}})})\text{ }\rangle :\text{ }=\dint \Psi (\lambda
_{1}^{(s_{1})},...,\lambda _{N}^{(s_{_{N}})})P_{(s_{1},...,s_{_{N}})}^{(%
\mathcal{E})}(\mathrm{d}\lambda _{1}^{(s_{1})}\times ...\times \mathrm{d}%
\lambda _{N}^{(s_{_{N}})})  \label{5}
\end{equation}%
the expected (mean) value of a bounded measurable real-valued function $\Psi
.$ In particular, 
\begin{equation}
\langle \text{ }\phi _{1}(\lambda _{1}^{(s_{1})})\cdot ...\cdot \phi
_{N}(\lambda _{N}^{(s_{_{N}})})\text{ }\rangle =\int \phi _{1}(\lambda
_{1}^{(s_{1})})\cdot ...\cdot \phi _{N}(\lambda
_{N}^{(s_{_{N}})})P_{(s_{1},...,s_{_{N}})}^{(\mathcal{E})}(\mathrm{d}\lambda
_{1}^{(s_{1})}\times ...\times \mathrm{d}\lambda _{N}^{(s_{_{N}})})
\label{6}
\end{equation}%
means the expectation of the product of bounded measurable real-valued
functions $\phi _{1}(\lambda _{1}^{(s_{1})}),$ $...,\phi _{N}(\lambda
_{N}^{(s_{_{N}})}).$ If outcomes observed at sites: $1\leq
n_{1}<...<n_{M}\leq N$, are real-valued and bounded then, for any $2\leq
M\leq N,$ the expectation of the product of outcomes observed at these
sites, that is:%
\begin{equation}
\langle \lambda _{n_{1}}^{(s_{n_{1}})}\cdot ...\cdot \lambda
_{n_{_{M}}}^{(s_{n_{_{M}}})}\rangle =\dint \lambda
_{n_{1}}^{(s_{n_{1}})}\cdot ...\cdot \lambda
_{n_{_{M}}}^{(s_{n_{_{M}}})}P_{(s_{1},...,s_{_{N}})}^{(\mathcal{E})}(\mathrm{%
d}\lambda _{1}^{(s_{1})}\times ...\times \mathrm{d}\lambda
_{N}^{(s_{_{N}})}),  \label{7}
\end{equation}%
is referred to as \emph{a} \emph{correlation function}. For $M=N$, this
correlation function is called \emph{full.}

If an $N$-partite joint measurement $(s_{1},...,s_{_{N}})\in \mathcal{E}$ is 
\emph{EPR\ local}\footnote{%
That is, local in the sense meant originally by Einstein, Podolsky and Rosen
in [21]. For details, see section 3 of [20].} then its probability
distribution and all marginals of this distribution depend only on settings
of the corresponding measurements at the corresponding sites, that is: $%
P_{(s_{1},...,s_{_{N}})}^{(\mathcal{E})}\equiv P_{(s_{1},...,s_{_{N}})}$ and 
\begin{eqnarray}
&&P_{(s_{1},...,s_{_{N}})}(\Lambda _{1}^{(s_{1})}\times ...\times \Lambda
_{_{n_{1}-1}}^{(s_{n_{1}-1})}\times \mathrm{d}\lambda
_{_{n_{1}}}^{(s_{n_{1}})}\times ...\times \mathrm{d}\lambda
_{_{n_{_{M}}}}^{(s_{n_{_{M}}})}\times \Lambda
_{n_{_{M}}+1}^{(s_{n_{_{M}}+1})}\times ...\times \Lambda _{_{N}}^{(s_{N})})
\label{3} \\
&\equiv &P_{(s_{n_{1}},...,s_{n_{_{M}}})}(\mathrm{d}\lambda
_{_{n_{1}}}^{(s_{n_{1}})}\times ...\times \mathrm{d}\lambda
_{_{n_{_{M}}}}^{(s_{n_{_{M}}})}),  \notag
\end{eqnarray}%
for any $1\leq n_{1}<...<n_{M}\leq N$ and any $1\leq M\leq N.$ In an EPR
local case, the probability distribution of outcomes observed by $n$-th
party under $s_{n}$-th measurement depends only on a setting of this
measurement and we denote it by%
\begin{equation}
P_{n}^{(s_{n})}(\mathrm{d}\lambda
_{n}^{(s_{n})}):=P_{(s_{1},...,s_{_{N}})}(\Lambda _{1}^{(s_{1})}\times
...\times \Lambda _{_{n-1}}^{(s_{n-1})}\times \mathrm{d}\lambda
_{n}^{(s_{n})}\times \Lambda _{_{n+1}}^{(s_{n+1})}\times ...\times \Lambda
_{_{N}}^{(s_{N})}).  \label{4}
\end{equation}

The main "qualitative" statements on a simulation of an $S_{1}\times
...\times S_{N}$-setting $N$-partite correlation experiment in terms of a
local hidden variable (LHV) model\footnote{%
For the definition of an LHV model, see section 4 of [20].} are introduced
in [20]. Below, we specify a single general representation for all linear
constraints, on either correlation functions or joint probabilities, arising
in the LHV frame. Particular cases of this general representation are
further considered in corollaries 1 - 4.

We stress that the EPR locality does not necessarily imply the existence for
a multipartite correlation experiment of an LHV model.

\begin{theorem}
Let an $S_{1}\times ...\times S_{N}$-setting $N$-partite correlation
experiment (\ref{1}), with outcomes $\lambda _{n}^{(s_{n})}\in \Lambda
_{n}^{(s_{n})},$ $s_{n}=1,...,S_{n},$ $n=1,...,N,$ of any spectral type,
discrete or continuous, admit an LHV model, conditional or unconditional.
Then the tight\footnote{%
The meaning of the term \emph{tight }is specified in\emph{\ }footnote 1. On
the difference between the terms \emph{tight} and \emph{extreme }with
respect to a linear LHV correlation constraint, see the end of section 2.1.}
linear unconditional LHV constraint on expectations:%
\begin{eqnarray}
&&\inf_{\lambda _{1}\in \Lambda _{1},...,\lambda _{_{N}}\in \Lambda _{_{N}}}%
\text{ }\dsum\limits_{s_{_{1}},...,s_{_{N}}}\Psi
_{(s_{1},...,s_{_{N}})}(\lambda _{1}^{(s_{_{1}})},...,\lambda
_{N}^{(s_{_{N}})})  \label{8} \\
&&  \notag \\
&\leq &\dsum\limits_{s_{_{1}},...,s_{_{N}}}\left\langle \Psi
_{(s_{1},...,s_{_{N}})}(\lambda _{1}^{(s_{_{1}})},...,\lambda
_{N}^{(s_{_{N}})})\right\rangle _{LHV}  \notag \\
&&  \notag \\
&\leq &\sup_{\lambda _{_{1}}\in \Lambda _{_{1}},...,\lambda _{_{N}}\in
\Lambda _{_{N}}}\text{ }\dsum\limits_{s_{_{1}},...,s_{_{N}}}\Psi
_{(s_{1},...,s_{_{N}})}(\lambda _{1}^{(s_{1})},...,\lambda
_{N}^{(s_{_{N}})}),  \notag
\end{eqnarray}%
holds for any collection $\{\Psi _{(s_{1},...,s_{_{N}})}\}$ of bounded
measurable real-valued functions, where $s_{_{1}}=1,...,S_{_{1}},$ $...,$ $%
s_{_{N}}=1,...,S_{_{N}},$ and $\lambda _{n}:=(\lambda _{n}^{(1)},...,\lambda
_{n}^{(S_{n})}),$ $\Lambda _{n}:=\Lambda _{n}^{(1)}\times ...\times \Lambda
_{n}^{(S_{n})}$.\newline
In particular, the tight linear LHV constraint on product expectations:%
\begin{eqnarray}
&&\inf_{\xi _{_{1}}\in \Phi _{_{1}},...,\xi _{N}\in \Phi _{N}}\text{ }%
F_{N}^{^{(\gamma )}}(\xi _{1},...,\xi _{N})  \label{9} \\
&&  \notag \\
&\leq &\dsum\limits_{s_{_{1}},...,s_{_{N}}}\gamma _{(s_{_{1}},...,s_{_{N}})}%
\text{ }\left\langle \phi _{1}^{(s_{1})}(\lambda _{1}^{(s_{1})})\cdot
...\cdot \phi _{N}^{(s_{_{N}})}(\lambda _{N}^{(s_{_{N}})})\right\rangle
_{LHV}  \notag \\
&&  \notag \\
&\leq &\sup_{\xi _{_{1}}\in \Phi _{_{1}},...,\xi _{N}\in \Phi _{N}}\text{ }%
F_{N}^{^{(\gamma )}}(\xi _{1},...,\xi _{N}),  \notag
\end{eqnarray}%
is valid for any bounded measurable real-valued functions $\phi
_{n}^{(s_{n})}(\lambda _{n}^{(s_{n})}),$ $\forall s_{n},$ $\forall n,$ and
any real coefficients $\gamma _{(s_{_{_{1}}},...,s_{_{_{N}}})}.$ Here,%
\begin{equation}
F_{N}^{^{(\gamma )}}(\xi _{1},...,\xi
_{N})=\dsum\limits_{s_{_{1}},...,s_{_{N}}}\text{ }\gamma
_{(s_{_{1}},...,s_{_{N}})}\text{ }\xi _{1}^{(s_{1})}\cdot ...\cdot \xi
_{N}^{(s_{_{N}})}  \label{10}
\end{equation}%
is an $N$-linear form of real vectors 
\begin{equation}
\xi _{n}=(\xi _{n}^{(1)},...,\xi _{n}^{(S_{n})})\in \mathbb{R}^{S_{n}},\text{
\ \ \ }n=1,...,N,
\end{equation}%
and, for any $n$ $\in \{1,...,N\},$%
\begin{eqnarray}
\Phi _{n} &=&\{\xi _{n}\in \mathbb{R}^{S_{n}}\mid \xi _{n}^{(s_{n})}=\phi
_{n}^{(s_{n})}(\lambda _{n}^{(s_{n})}),\text{ \ \ }\lambda _{n}^{(s_{n})}\in
\Lambda _{n}^{(s_{n})},\text{ \ }s_{n}=1,...,S_{n}\}  \label{11} \\
&\subset &\mathbb{R}^{S_{n}}  \notag
\end{eqnarray}%
is the range of the bounded vector-valued function with components $\phi
_{n}^{(s_{n})}(\lambda _{n}^{(s_{n})}).\medskip $
\end{theorem}

\begin{proof}
In view of (\ref{5}),%
\begin{eqnarray}
&&\dsum\limits_{s_{_{1}},...,s_{_{N}}}\left\langle \Psi
_{(s_{_{1}},...,s_{_{N}})}(\lambda _{1}^{(s_{_{1}})},...,\lambda
_{N}^{(s_{_{N}})})\right\rangle  \label{12} \\
&=&\dsum\limits_{s_{_{1}},...,s_{_{N}}}\int \Psi
_{(s_{_{1}},...,s_{_{N}})}(\lambda _{1}^{(s_{_{1}})},...,\lambda
_{N}^{(s_{_{N}})})P_{(s_{1},...,s_{_{N}})}^{(\mathcal{E)}}(\mathrm{d}\lambda
_{1}^{(s_{1})}\times ...\times \mathrm{d}\lambda _{N}^{(s_{_{N}})}).  \notag
\end{eqnarray}%
Let family (\ref{1}) admit an LHV model. Then, by statement \textrm{(c)} of
theorem 1 in [20], there exists a joint probability measure%
\begin{equation}
\mu _{\mathcal{E}}(\mathrm{d}\lambda _{1}^{(1)}\times ...\times \mathrm{d}%
\lambda _{1}^{(S_{1})}\times ...\times \mathrm{d}\lambda _{N}^{(1)}\times
...\times \mathrm{d}\lambda _{N}^{(S_{N})})  \label{13}
\end{equation}%
of all outcomes observed at all sites that returns each distribution $%
P_{(s_{1},...,s_{_{N}})}^{(\mathcal{E)}}$ of family (\ref{1}) as the
corresponding marginal. Taking this property into account in relation (\ref%
{12}), we have:%
\begin{eqnarray}
&&\dsum\limits_{s_{_{1}},...,s_{_{N}}}\left\langle \Psi
_{(s_{1},...,s_{_{N}})}(\lambda _{1}^{(s_{_{1}})},...,\lambda
_{N}^{(s_{_{N}})})\right\rangle _{LHV}  \label{14} \\
&=&\dint \{\dsum\limits_{s_{_{1}},...,s_{_{N}}}\Psi
_{(s_{1},...,s_{_{N}})}(\lambda _{1}^{(s_{1})},...,\lambda
_{N}^{(s_{_{N}})})\}\text{ }\mu _{\mathcal{E}}(\mathrm{d}\lambda _{1}\times
...\times \mathrm{d}\lambda _{N}),  \notag
\end{eqnarray}%
where, for short, we denote $\lambda _{n}=(\lambda _{n}^{(1)},...,\lambda
_{n}^{(S_{n})})$ and $\Lambda _{n}=\Lambda _{n}^{(1)}\times ...\times
\Lambda _{n}^{(S_{n})}.$ Considering the least upper bound of the second
line in (\ref{14}), we derive:%
\begin{eqnarray}
&&\dsum\limits_{s_{_{1}},...,s_{_{N}}}\left\langle \Psi
_{(s_{1},...,s_{_{N}})}(\lambda _{1}^{(s_{_{1}})},...,\lambda
_{N}^{(s_{_{N}})})\right\rangle _{LHV}  \label{15} \\
&\leq &\sup_{\lambda _{1}\in \Lambda _{1},...,\lambda _{_{N}}\in \Lambda
_{_{N}}}\text{ }\dsum\limits_{s_{_{1}},...,s_{_{N}}}\Psi
_{(s_{1},...,s_{_{N}})}(\lambda _{1}^{(s_{1})},...,\lambda
_{N}^{(s_{_{N}})}).  \notag
\end{eqnarray}%
The left-hand side bound of (\ref{8}) is proved quite similarly.

In order to prove (\ref{9}), let us specify (\ref{8}) with functions $\Psi
_{(s_{_{1}},...,s_{_{N}})}$ of the product form:%
\begin{equation}
\Psi _{(s_{1},...,s_{_{N}})}(\lambda _{1}^{(s_{1})},...,\lambda
_{N}^{(s_{_{N}})})=\gamma _{(s_{_{1}},...,s_{_{N}})}\text{ }\phi
_{1}^{(s_{1})}(\lambda _{1}^{(s_{1})})\cdot ...\cdot \phi
_{N}^{(s_{_{N}})}(\lambda _{N}^{(s_{_{N}})}).  \label{16}
\end{equation}%
For these functions,%
\begin{eqnarray}
&&\sup_{\lambda _{1}\in \Lambda _{1},...,\lambda _{N}\in \Lambda _{N}}\text{ 
}\dsum\limits_{s_{_{1}},...,s_{_{N}}}\Psi _{(s_{_{1}},...,s_{_{N}})}(\lambda
_{1}^{(s_{1})},...,\lambda _{N}^{(s_{_{N}})})  \label{17} \\
&&  \notag \\
&=&\sup_{\lambda _{1}\in \Lambda _{1},...,\lambda _{_{N}}\in \Lambda _{_{N}}}%
\text{ }\dsum\limits_{s_{_{1}},,...,s_{_{N}}}\text{ }\gamma
_{(s_{_{1}},...,s_{_{N}})}\text{ }\phi _{1}^{(s_{1})}(\lambda
_{1}^{(s_{1})})\cdot ...\cdot \phi _{N}^{(s_{_{N}})}(\lambda
_{N}^{(s_{_{N}})}).  \notag
\end{eqnarray}%
Denoting $\xi _{n}^{(s_{n})}=\phi _{n}^{(s_{n})}(\lambda _{n}^{(s_{n})})$
and taking into account (\ref{10}) - (\ref{11}), we have:%
\begin{eqnarray}
&&\sup_{\lambda _{1}\in \Lambda _{1},...,\lambda _{N}\in \Lambda _{N}}\text{ 
}\sum \gamma _{(s_{_{1}},...,s_{_{N}})}\text{ }\phi _{1}^{(s_{1})}(\lambda
_{1}^{(s_{1})})\cdot ...\cdot \phi _{N}^{(s_{_{N}})}(\lambda
_{N}^{(s_{_{N}})})  \label{18} \\
&&  \notag \\
&=&\sup_{\xi _{_{1}}\in \Phi _{_{1}},...,\xi _{N}\in \Phi _{N}}\text{ }%
F_{N}^{^{(\gamma )}}(\xi _{1},...,\xi _{N}).  \notag
\end{eqnarray}%
$\medskip $The left-hand side of (\ref{9}) is proved quite similarly.
\end{proof}

\bigskip

If an $S_{1}\times ...\times S_{N}$-setting $N$-partite correlation
experiment (\ref{1}) admits a \emph{conditional} LHV model then linear
combinations of expectations satisfy not only unconditional LHV constraints (%
\ref{8}), (\ref{9}) but also their conditional versions - with the
corresponding \emph{conditional }supremums and infimums.\emph{\ }The LHV
model considered by Bell in [1] represents an example of a conditional LHV
model.

Depending on a choice of functions, standing in (\ref{9}), this constraint
reduces to either a general representation for all LHV constraints on
correlation functions or a general representation for all LHV constraints on
joint probabilities.

\subsection{Constraints on correlation functions}

Consider an $S_{1}\times ...\times S_{N}$-setting $N$-partite correlation
experiment with real-valued outcomes $\lambda _{n}^{(s_{n})}\in \Lambda
_{n}^{(s_{n})}\subseteq \lbrack -1,1]$ of any spectral type, discrete or
continuous, such that%
\begin{equation}
\sup \Lambda _{n}^{(s_{n})}=1,\text{ \ \ \ }\inf \Lambda _{n}^{(s_{n})}=-1,%
\text{ \ \ \ }\forall s_{n},\text{ }\forall n.  \label{19'}
\end{equation}%
Note that the description of any multipartite correlation experiment, with
at least two outcomes at each site, can be reduced to this case.

For this correlation experiment, let us specify the LHV constraint (\ref{9})
with functions 
\begin{equation}
\phi _{n}^{(s_{n})}(\lambda _{n}^{(s_{n})})=\lambda
_{n}^{(s_{n})}+z_{n}^{(s_{n})},\text{ \ \ \ \ }\forall s_{n},\text{ }\forall
n,  \label{19}
\end{equation}%
where each $z_{n}^{(s_{n})}$ is an arbitrary real number. We derive:%
\begin{eqnarray}
&&\dsum\limits_{s_{_{1}},...,s_{_{N}}}\gamma _{(s_{_{1}},...,s_{_{N}})}\text{
}\left\langle \phi _{1}^{(s_{1})}(\lambda _{1}^{(s_{1})})\cdot ...\cdot \phi
_{N}^{(s_{_{N}})}(\lambda _{N}^{(s_{_{N}})})\right\rangle  \label{20} \\
&=&\dsum\limits_{s_{_{1}},...,s_{_{N}}}\gamma _{(s_{_{1}},...,s_{_{N}})}%
\text{ }z_{1}^{(s_{1})}\cdot ...\cdot z_{N}^{(s_{N})}  \notag \\
&&+\sum_{\substack{ 1\leq n_{1}<...<n_{M}\leq N,  \\ M=1,...,N}}\text{ }%
\sum_{s_{_{n_{1}}},...,s_{_{n_{_{M}}}}}\gamma
_{(s_{_{n_{1}}},...,s_{n_{_{M}}})}\text{ }\left\langle \lambda
_{n_{1}}^{(s_{_{n_{1}}})}\cdot ...\cdot \lambda
_{n_{_{M}}}^{(s_{_{n_{_{M}}}})}\right\rangle ,  \notag
\end{eqnarray}%
where\footnote{%
Here, $\delta _{M,N}=1$ if $M=N$ and $\delta _{M,N}=0$ if $M\neq N$.}%
\begin{eqnarray}
\gamma _{(s_{_{n_{1}}},...,s_{n_{_{M}}})} &:&=\gamma
_{(s_{_{1}},...,s_{_{N}})}\text{ }\delta _{M,N}+  \label{21} \\
&&+(1-\delta _{M,N})\dsum\limits_{s_{n},\forall n\neq
n_{_{1}},...,n_{_{M}}}\{\gamma
_{(s_{1},...,s_{_{n}},...,s_{_{N}})}\tprod\limits_{n\neq
n_{_{1}},...,n_{_{M}}}z_{n}^{(s_{n})}\}.  \notag
\end{eqnarray}

For the $N$-linear form (\ref{10}), consider the change of variables: $\xi
_{n}=\eta _{n}+z_{n},$ $\forall n\in \{1,...,N\},$ where $%
z_{n}:=(z_{n}^{(1)},...,z_{n}^{(S_{n})})\in \mathbb{R}^{S_{n}}$ is the real
vector with components given by real numbers in (\ref{19}). We have:%
\begin{eqnarray}
F_{N}^{(\gamma )}(\xi _{1},...,\xi _{N})
&=&\dsum\limits_{s_{_{1}},...,s_{_{N}}}\gamma _{(s_{_{1}},...,s_{_{N}})}%
\text{ }z_{1}^{(s_{1})}\cdot ...\cdot z_{N}^{(s_{_{N}})}  \label{22} \\
&&+\sum_{\substack{ 1\leq n_{1}<...<n_{M}\leq N,  \\ M=1,...,N}}\text{ }%
\sum_{s_{_{n_{1}}},...,s_{_{n_{_{M}}}}}\gamma
_{(s_{_{n_{1}}},...,s_{n_{_{M}}})}\text{ }\eta _{n_{1}}^{(s_{n_{1}})}\cdot
...\cdot \eta _{n_{_{M}}}^{(s_{_{n_{_{M}}}})}.  \notag
\end{eqnarray}%
\medskip From (\ref{11}), (\ref{19}) it follows that%
\begin{equation}
\xi _{n}\in \Phi _{n}\text{ \ \ }\Leftrightarrow \text{ \ \ }\eta _{n}\in
\Lambda _{n}=\Lambda _{n}^{(1)}\times ...\times \Lambda
_{n}^{(S_{n})}\subseteq \mathbb{[}-1,1]^{S_{n}},  \label{22'}
\end{equation}%
where, due to (\ref{19'}), closure $\overline{\Lambda }_{n}$ of the bounded
set $\Lambda _{n}$ satisfies the relation $\{-1,1\}^{S_{n}}\subseteq 
\overline{\Lambda }_{n}\subseteq \mathbb{[}-1,1]^{S_{n}}.$

Substituting (\ref{20}), (\ref{22}) into (\ref{9}) and taking into account (%
\ref{22'}), we derive:%
\begin{eqnarray}
&&\inf_{\eta _{_{1}}\in \Lambda _{1},...,\eta _{_{N}}\in \Lambda _{_{N}}}%
\text{ }\sum_{\substack{ 1\leq n_{1}<...<n_{M}\leq N,  \\ M=1,...,N}}%
F_{M}^{(\gamma )}(\eta _{n_{1}},...,\eta _{n_{M}})  \label{23} \\
&&  \notag \\
&\leq &\sum_{\substack{ 1\leq n_{1}<...<n_{M}\leq N,  \\ M=1,...,N}}\text{ \ 
}\sum_{s_{_{n_{1}}},...,s_{_{n_{_{M}}}}}\gamma
_{(s_{_{n_{1}}},...,s_{n_{_{M}}})}\text{ }\left\langle \lambda
_{n_{1}}^{(s_{_{n_{1}}})}\cdot ...\cdot \lambda
_{n_{_{M}}}^{(s_{_{n_{_{M}}}})}\right\rangle _{LHV}  \notag \\
&&  \notag \\
&\leq &\sup_{\eta _{_{1}}\in \Lambda _{1},...,\eta _{_{N}}\in \Lambda
_{_{N}}}\text{ }\sum_{\substack{ 1\leq n_{1}<...<n_{M}\leq N,  \\ M=1,...,N}}%
F_{M}^{(\gamma )}(\eta _{n_{1}},...,\eta _{n_{M}}),  \notag
\end{eqnarray}%
where 
\begin{equation}
F_{M}^{(\gamma )}(\eta _{n_{1}},...,\eta
_{n_{_{M}}})=\sum_{s_{_{n_{1}}},...,s_{_{n_{_{M}}}}}\gamma
_{_{(s_{_{n_{1}}},...,s_{n_{_{M}}})}}\text{ }\eta
_{n_{1}}^{(s_{n_{1}})}\cdot ...\cdot \eta _{n_{_{M}}}^{(s_{_{n_{_{M}}}})}
\label{24}
\end{equation}%
is an $M$-linear form of real vectors $\eta _{1}=(\eta _{1}^{(1)},...,\eta
_{1}^{(S_{1})})\in \mathbb{R}^{S_{1}},$ $...,$ $\eta _{N}=(\eta
_{N}^{(1)},...,\eta _{N}^{(S_{N})})\in \mathbb{R}^{S_{N}}.$

For a further simplification of constraint (\ref{23}), we need the following
property proved in appendix.

\begin{lemma}
Let, for each bounded set $\Lambda _{n}\subseteq \lbrack -1,1]^{S_{n}},$ $%
n\in \{1,...,N\},$ its closure $\overline{\Lambda }_{n}$ satisfies the
relation: 
\begin{equation}
\{-1,1\}^{S_{n}}\subseteq \overline{\Lambda }_{n}\subseteq \lbrack
-1,1]^{S_{n}},\text{ \ \ \ }\forall n=1,...,N.  \label{25'}
\end{equation}%
Then 
\begin{eqnarray}
&&\sup_{\eta _{_{1}}\in \Lambda _{1},...,\eta _{_{N}}\in \Lambda _{_{N}}}%
\text{ }\sum_{\substack{ 1\leq n_{_{1}}<...<n_{M}\leq N,  \\ M=1,...,N}}%
F_{M}^{^{(\gamma )}}(\eta _{n_{_{1}}},...,\eta _{n_{_{M}}})  \label{25} \\
&&  \notag \\
&=&\max_{\eta _{_{1}}\in \{-1,1\}^{S_{1}},...,\eta _{_{N}}\in
\{-1,1\}^{S_{N}}}\text{\ }\sum_{\substack{ 1\leq n_{_{1}}<...<n_{M}\leq N, 
\\ M=1,...,N}}F_{M}^{^{(\gamma )}}(\eta _{n_{_{1}}},...,\eta _{n_{_{M}}}), 
\notag
\end{eqnarray}%
with a similar expression for infimum.
\end{lemma}

\bigskip

Substituting (\ref{25}) into constraint (\ref{23}), we derive the following
corollary of theorem 1.

\begin{corollary}
Let an $S_{1}\times ...\times S_{N}$-setting $N$-partite correlation
experiment (\ref{1}), with real-valued outcomes 
\begin{equation}
\lambda _{n}^{(s_{n})}\in \Lambda _{n}^{(s_{n})}\subseteq \lbrack -1,1],%
\text{ \ \ \ \ }\sup \Lambda _{n}^{(s_{n})}=1,\text{ \ \ }\inf \Lambda
_{n}^{(s_{n})}=-1,\text{ \ \ \ }\forall s_{n},\text{ }\forall n,
\end{equation}%
of any spectral type, discrete or continuous, admit an LHV model. Then the
tight linear LHV constraint on correlation functions:%
\begin{eqnarray}
&&\min_{\eta _{_{1}}\in \{-1,1\}^{S_{1}},...,\eta _{_{N}}\in
\{-1,1\}^{S_{N}}}\text{ }\sum_{\substack{ 1\leq n_{1}<...<n_{M}\leq N,  \\ %
M=1,...,N}}F_{M}^{^{(\gamma )}}(\eta _{n_{_{1}}},...,\eta _{n_{_{M}}})
\label{26} \\
&&  \notag \\
&&  \notag \\
&\leq &\sum_{\substack{ 1\leq n_{1}<...<n_{M}\leq N,  \\ M=1,...,N}}\text{ \ 
}\sum_{s_{_{n_{1}}},...,s_{_{n_{_{M}}}}}\gamma
_{(s_{_{n_{_{1}}}},...,s_{n_{_{M}}})}\text{ }\left\langle \lambda
_{n_{_{1}}}^{(s_{_{n_{1}}})}\cdot ...\cdot \lambda
_{n_{_{M}}}^{(s_{_{n_{_{M}}}})}\right\rangle _{LHV}  \notag \\
&&  \notag \\
&&  \notag \\
&\leq &\max_{\eta _{_{1}}\in \{-1,1\}^{S_{1}},...,\eta _{_{N}}\in
\{-1,1\}^{S_{N}}}\text{ }\sum_{\substack{ 1\leq n_{1}<...<n_{M}\leq N,  \\ %
M=1,...,N}}F_{M}^{^{(\gamma )}}(\eta _{n_{_{1}}},...,\eta _{n_{_{M}}}), 
\notag
\end{eqnarray}%
$\bigskip $holds for any collection $\{\gamma
_{_{(s_{_{n_{1}}},...,s_{n_{_{M}}})}}\}$ of real coefficients. Here, $%
F_{M}^{(\gamma )}$ is an $M$-linear form defined by (\ref{24}) and extremums
are taken over all $2^{S_{1}+...+S_{N}}$ vertices of hypercube $%
[-1,1]^{S_{1}+...+S_{N}}$ $\subset \mathbb{R}^{S_{1}+...+S_{N}}.$
\end{corollary}

\bigskip

From the definition of a Bell-type inequality, given in introduction, and
corollary 1 it follows that \emph{the form of} \emph{any correlation
Bell-type inequality does not depend on a spectral type of outcomes observed
at each site, in particular, on their number and is determined only by
extremal values of these outcomes.}

If, in particular, $\gamma _{(s_{_{n_{1}}},...,s_{n_{_{M}}})}=\delta _{N,M}$ 
$\gamma _{(s_{1},...,s_{N})},$ then (\ref{26}) reduces to the tight linear
LHV constraint on the full correlation functions: 
\begin{eqnarray}
&&\min_{(\eta _{_{1}},...,\eta _{_{N}})\in \{-1,1\}^{d}}\text{ }%
F_{N}^{^{(\gamma )}}(\eta _{1},...,\eta _{N})\text{ }  \label{27} \\
&&  \notag \\
&\leq &\sum_{s_{1},...,s_{_{N}}}\gamma _{(s_{_{1}},...,s_{_{N}})}\text{ }%
\langle \lambda _{1}^{(s_{_{1}})}\cdot ...\cdot \lambda
_{N}^{(s_{_{N}})}\rangle _{LHV}  \notag \\
&&  \notag \\
&\leq &\max_{(\eta _{_{1}},...,\eta _{_{N}})\in \{-1,1\}^{d}}\text{ }%
F_{N}^{^{(\gamma )}}(\eta _{1},...,\eta _{N}),  \notag
\end{eqnarray}%
where $d:=S_{1}+...+S_{N}.$ Noting that 
\begin{equation}
F_{N}^{^{(\gamma )}}(\eta _{1},...,\eta _{n},...,\eta
_{N})=-F_{N}^{^{(\gamma )}}(\eta _{1},...,-\eta _{n},...,\eta _{N}),
\label{28}
\end{equation}%
and points%
\begin{equation}
(\eta _{1},...,\eta _{n},...,\eta _{N})\in \mathbb{R}^{d},\text{ \ \ \ \ }%
(\eta _{1},...,-\eta _{n},...,\eta _{N})\in \mathbb{R}^{d}  \label{29}
\end{equation}%
belong to hypercube $[-1,1]^{d}\subset \mathbb{R}^{d}$ simultaneously, we
derive:%
\begin{eqnarray}
-\min_{(\eta _{_{1}},...,\eta _{_{N}})\text{ }\in \text{ }\{-1,1\}^{d}}\text{
}F_{N}^{^{(\gamma )}}(\eta _{1},...,\eta _{N}) &=&\max_{(\eta
_{_{1}},...,\eta _{_{N}})\text{ }\in \text{ }\{-1,1\}^{d}}\text{ }%
F_{N}^{^{(\gamma )}}(\eta _{1},...,\eta _{N})  \label{30} \\
&&  \notag \\
&=&\max_{(\eta _{_{1}},...,\eta _{_{N}})\text{ }\in \text{ }\{-1,1\}^{d}}%
\text{ }\left\vert F_{N}^{^{(\gamma )}}(\eta _{1},...,\eta _{N})\right\vert .
\notag
\end{eqnarray}%
Substituting (\ref{30}) into (\ref{27}), we come to the following corollary
of theorem 1.

\begin{corollary}
Let an $S_{1}\times ...\times S_{N}$-setting $N$-partite correlation
experiment (\ref{1}), with real-valued outcomes $\lambda _{n}^{(s_{n})}\in
\Lambda _{n}^{(s_{n})}\subseteq \lbrack -1,1],$ $\sup \Lambda
_{n}^{(s_{n})}=1,$ $\inf \Lambda _{n}^{(s_{n})}=-1,\ \forall s_{n},\forall
n, $ of any spectral type, discrete or continuous, admit an LHV model. Then
the full correlation functions satisfy the tight linear LHV constraint%
\begin{equation}
\left\vert \sum_{s_{_{_{1}}},...,s_{_{_{N}}}}\gamma _{(s_{1},...,s_{N})}%
\text{ }\left\langle \lambda _{1}^{(s_{_{1}})}\cdot ...\cdot \lambda
_{N}^{(s_{_{N}})}\right\rangle _{LHV}\right\vert \text{ }\leq \text{ }\max 
_{\substack{ \eta _{_{1}}\in \{-1,1\}^{S_{1}},...,  \\ \eta _{_{N}}\in
\{-1,1\}^{S_{N}}}}\text{ }\left\vert F_{N}^{^{(\gamma )}}(\eta _{1},...,\eta
_{N})\right\vert ,  \label{31}
\end{equation}%
for any real coefficients $\gamma _{(s_{1},...,s_{N})}.$
\end{corollary}

\medskip

If a correlation experiment admits a conditional LHV model then, in addition
to (\ref{26}), (\ref{31}), the correlation functions satisfy also the
conditional versions of these constraints - with the corresponding
conditional extremums. The original Bell inequality, derived by Bell in [1]
in the frame of the conditional LHV model, represents an example of a
conditional LHV constraint on the full correlation functions.

We stress that, in corollaries 1, 2, the term \emph{a tight linear LHV
constraint} does not mean \emph{an extreme linear LHV constraint. }The
difference between these two terms is clearly seen due to the geometric
interpretation of, say, constraint (\ref{31}) in terms of the polytope
approach [19].

Namely, for any choice of coefficients $\gamma _{(s_{_{1}},...,s_{_{N}})}$
in constraint (\ref{31}) represented otherwise as: 
\begin{eqnarray}
-\max_{(\eta _{_{1}},...,\eta _{_{N}})\in \{-1,1\}^{d}}\text{ }\left\vert
F_{N}^{^{(\gamma )}}(\eta _{1},...,\eta _{N})\right\vert &\leq
&\sum_{s_{_{1}},...,s_{_{N}}}\gamma _{(s_{_{1}},...,s_{_{N}})}\text{ }%
\left\langle \lambda _{1}^{(s_{_{1}})}\cdot ...\cdot \lambda
_{N}^{(s_{_{N}})}\right\rangle _{LHV}\text{ }  \label{32} \\
&&  \notag \\
&\leq &\max_{(\eta _{_{1}},...,\eta _{_{N}})\in \{-1,1\}^{d}}\text{ }%
\left\vert F_{N}^{^{(\gamma )}}(\eta _{1},...,\eta _{N})\right\vert ,  \notag
\end{eqnarray}%
the right-hand side (or the left-hand side) inequality describes the half
space, defined by the hyperplane passing \emph{outside} of the corresponding
polytope via at least one of its vertices. A tight linear LHV inequality
becomes an extreme one whenever this hyperplane describes a face of the
corresponding polytope.

\subsection{Constraints on joint probabilities}

For an $S_{1}\times ...\times S_{N}$-setting $N$-partite correlation, with
at least $Q_{n}+1$ (possibly, infinitely many) outcomes at each site, let us
specify constraint (\ref{9}) with functions\footnote{%
Here, $\chi _{_{D}}(\lambda ),$ $\lambda \in \Lambda ,$ is an indicator
function of a subset $D\subseteq \Lambda ,$ defined by relations: $\chi
_{D}(\lambda )=1$ if $\lambda \in D$ and $\chi _{D}(\lambda )=0\ $if\ $%
\lambda \notin D.$}%
\begin{equation}
\phi _{n}^{(s_{n})}(\lambda _{n}^{(s_{n})})=\sum_{q_{n}=1,...,Q_{n}}\{\tau
_{n}^{(s_{n},q_{n})}\chi _{D_{n}^{(s_{n},q_{n})}}(\lambda
_{n}^{(s_{n})})+z_{n}^{(s_{n},q_{n})}\},  \label{33}
\end{equation}%
where $\tau _{n}^{(s_{n},q_{n})}$ and $z_{n}^{(s_{n},q_{n})}$ are arbitrary
real numbers and $D_{n}^{(s_{n},q_{n})}\subset \Lambda _{n}^{(s_{n})}$, $%
D_{n}^{(s_{n},q_{n})}\neq \varnothing ,$ $q_{n}\in \{1,...,Q_{n}\},$ are any
mutually disjoint subsets: $D_{n}^{(s_{n},q_{n})}\cap
D_{n}^{(s_{n},q_{n}^{\prime })}=\varnothing ,$ $\forall q_{n}\neq
q_{n}^{\prime },$ observed under $s_{n}$-th measurement at $n$-th site and
such that $\cup _{q_{n}}D_{n}^{(s_{n},q_{n})}\neq \Lambda _{n}^{(s_{n})}.$

Substituting these functions into the LHV constraint (\ref{9}), making
transformations similar to those in section 2.1 and renaming coefficients,
we come to following corollary of theorem 1.

\begin{corollary}
Let an $S_{1}\times ...\times S_{N}$-setting $N$-partite correlation
experiment (\ref{1}), satisfying the EPR locality\footnote{%
See condition (\ref{3}) and notation (\ref{4}).} and with at least $%
(Q_{n}+1) $ outcomes at each $n$-th site, admit an LHV model. Then the tight
linear LHV constraint on joint probabilities:%
\begin{eqnarray}
&&\min_{\eta _{_{1}}\in \Xi _{_{1}},...,\eta _{_{N}}\in \Xi _{N}}\text{\ }%
\tsum\limits_{\substack{ 1\leq n_{_{1}},...,n_{_{M}}\leq N,  \\ M=1,...,N}}%
F_{M}^{^{(\gamma )}}(\eta _{n_{_{1}}},...,\eta _{n_{_{M}}})  \label{34} \\
&&  \notag \\
&\leq &\text{ \ }\sum_{\substack{ 1\leq n_{_{1}}<...<n_{_{M}}\leq N,  \\ %
M=1,...,N}}\text{ }\sum_{\substack{ s_{n_{_{1}}},...,s_{n_{_{M}}},  \\ %
q_{n_{_{_{1}}}},...,q_{n_{_{M}}}}}\gamma
_{(s_{n_{_{1}}},...,s_{n_{_{M}}})}^{^{(q_{n_{_{_{1}}}},...,q_{n_{_{M}}})}}P_{(s_{_{n_{_{1}}}},...,s_{_{n_{_{M}}}})}(D_{n_{1}}^{(s_{n_{_{1}}},q_{n_{1}})}\times ...\times D_{n_{_{M}}}^{(s_{_{n_{_{M}}}},q_{_{n_{M}}})})
\notag \\
&&  \notag \\
&\leq &\text{\ }\max_{\eta _{_{1}}\in \text{ }\Xi _{_{1}},...,\eta
_{_{N}}\in \Xi _{N}}\text{\ }\tsum\limits_{\substack{ 1\leq
n_{1},...,n_{M}\leq N,  \\ M=1,...,N}}F_{M}^{^{(\gamma )}}(\eta
_{n_{_{1}}},...,\eta _{n_{_{M}}}),  \notag
\end{eqnarray}%
\bigskip holds for an arbitrary collection $\{\gamma
_{(s_{n_{1}},...,s_{n_{_{M}}})}^{(q_{n_{1}},...,q_{n_{_{M}}})}\}$ of real
coefficients and any events $D_{n}^{(s_{n},q_{n})}\subset \Lambda
_{n}^{(s_{n})},$ $D_{n}^{(s_{n},q_{n})}\neq \varnothing ,$ $%
q_{n}=1,...,Q_{n},$ observed under $s_{n}$-th measurement at an $n$-th site,
such that, for any $Q_{n}\geq 2,$ these events are mutually incompatible: $%
D_{n}^{(s_{n},q_{n})}\cap D_{n}^{(s_{n},q_{n}^{\prime })}=\varnothing ,$ $%
\forall q_{n}\neq q_{n}^{\prime },$ and satisfy the relation%
\begin{equation}
\cup _{q_{n}=1,...,Q_{n}}D_{n}^{(s_{n},q_{n})}\neq \Lambda _{n}^{(s_{n})}.
\label{35}
\end{equation}%
In (\ref{34}),\ 
\begin{equation}
F_{M}^{^{(\gamma )}}(\eta _{n_{1}},...,\eta _{n_{_{M}}})=\sum_{_{\substack{ %
s_{n_{_{1}}},...,s_{n_{_{M}}},  \\ q_{n_{_{_{1}}}},...,q_{_{n_{M}}}}}}\gamma
_{(s_{_{n_{_{1}}}},...,s_{n_{_{M}}})}^{(q_{_{n_{_{1}}}},...,q_{_{n_{_{M}}}})}%
\text{ }\eta _{n_{_{1}}}^{(s_{n_{1}},q_{n_{1}})}\cdot ...\cdot \eta
_{n_{_{M}}}^{(s_{_{n_{_{M}}}},q_{_{n_{_{M}}}})}  \label{36}
\end{equation}%
is an $M$-linear form of real vectors $\eta _{n}\in \mathbb{R}^{S_{n}Q_{n}}$%
, with components $\eta _{n}^{(s_{n},q_{n})},$ and\medskip 
\begin{equation}
\Xi _{n}=\{\eta _{n}\in \{0,1\}^{S_{n}Q_{n}}\mid
\sum_{q_{n}=1,...,Q_{n}}\eta _{n}^{(s_{n},q_{n})}\in \{0,1\},\text{ \ }%
\forall s_{n}=1,...,S_{n}\},  \label{37}
\end{equation}%
for any $n=1,...,N.$
\end{corollary}

\bigskip

For an $S_{1}\times S_{2}$-setting bipartite correlation experiment, the LHV
constraint (\ref{34}) takes the form:%
\begin{eqnarray}
&&\min_{\eta _{1}\in \text{ }\Xi _{1},\text{ }\eta _{_{2}}\in \Xi
_{2}}\left\{ F_{2}^{(\gamma )}(\eta _{1},\eta _{2})+F_{1}^{(\gamma
_{1})}(\eta _{1})+F_{1}^{(\gamma _{2})}(\eta _{2})\right\}  \label{38} \\
&&  \notag \\
&\leq &\dsum\limits_{\substack{ s_{1},s_{2},  \\ q_{1},q_{2}}}\gamma
_{_{(s_{1},s_{2})}}^{^{(q_{1},q_{2})}}P_{(s_{1},s_{2})}(D_{1}^{^{(s_{1},q_{1})}}\times D_{2}^{(s_{_{2}},q_{_{2}})})%
\text{ }+\text{ }\tsum_{s_{1},q_{1}}\gamma
_{1}^{^{(s_{1},q_{1})}}P_{1}^{(s_{1})}(D_{1}^{(s_{1},q_{1})})  \notag \\
&&  \notag \\
&&+\tsum_{s_{2},q_{2}}\gamma
_{2}^{^{(s_{2},q_{2})}}P_{2}^{(s_{2})}(D_{2}^{(s_{_{2}},q_{_{2}})})  \notag
\\
&&  \notag \\
&\leq &\max_{\eta _{1}\in \text{ }\Xi _{1},\text{ }\eta _{_{2}}\in \Xi
_{2}}\left\{ F_{2}^{(\gamma )}(\eta _{1},\eta _{2})+F_{1}^{(\gamma
_{1})}(\eta _{1})+F_{1}^{(\gamma _{2})}(\eta _{2})\right\} ,  \notag
\end{eqnarray}%
where: (i) $\gamma =(\gamma _{_{(s_{1},s_{2})}}^{^{(q_{1},q_{2})}})$ is a
real matrix of dimension $S_{1}Q_{1}\times S_{2}Q_{2};$ (ii) $\gamma _{1}\in 
\mathbb{R}^{S_{1}Q_{1}},$ $\gamma _{2}\in \mathbb{R}^{S_{2}Q_{2}}$ are any
real vectors with components $\gamma _{1}^{(s_{1},q_{1})}$, $\gamma
_{2}^{(s_{2},q_{2})};$ (iii) $F_{2}^{(\gamma )}$is a bilinear form and $%
F_{1}^{(\gamma _{1})},$ $F_{1}^{(\gamma _{2})}$ are \textrm{1}-linear forms,
given by:%
\begin{eqnarray}
F_{2}^{(\gamma )}(\eta _{1},\eta _{2}) &=&\dsum\limits_{\substack{ %
s_{1},s_{2},  \\ q_{1},q_{2}}}\gamma _{_{(s_{1},s_{2})}}^{^{(q_{1},q_{2})}}%
\text{ }\eta _{1}^{(s_{1},q_{1})}\eta _{2}^{(s_{2},q_{2})}=(\eta _{1},\gamma
\eta _{2}),  \label{39} \\
&&  \notag \\
F_{1}^{(\gamma _{1})}(\eta _{1}) &=&\tsum_{s_{1},q_{1}}\gamma
_{1}^{(s_{1},q_{1})}\eta _{1}^{(s_{1},q_{1})}=(\eta _{1},\gamma _{1}), 
\notag \\
&&  \notag \\
F_{1}^{(\gamma _{2})}(\eta _{2}) &=&\tsum_{s_{2},q_{2}}\gamma
_{2}^{(s_{2},q_{2})}\eta _{2}^{(s_{2},q_{2})}=(\eta _{2},\gamma _{2}). 
\notag
\end{eqnarray}%
\smallskip \smallskip Here, $(\cdot ,\cdot )$ denotes the scalar product on
the corresponding space $\mathbb{R}^{SQ}$.

Finally, let us specify the general form of tight LHV constraints on joint
probabilities of \emph{arbitrary} events, not necessarily of the product
form. Taking in constraint (\ref{8}) functions%
\begin{equation}
\Psi _{s}(\lambda _{1}^{(s_{_{1}})},...,\lambda
_{N}^{(s_{_{N}})})=\sum_{q_{s}}\gamma _{_{s}}^{^{(q_{s})}}\chi
_{_{D_{_{s}}^{(q_{s})}}}(\lambda _{1}^{(s_{_{1}})},...,\lambda
_{N}^{(s_{_{N}})}),
\end{equation}%
where $D_{_{s}}^{(q_{s})}\subseteq \Lambda _{1}^{(s_{1})}\times ...\times
\Lambda _{N}^{(s_{_{N}})},$ $q_{s}=1,...,Q_{s},$ are any events observed
under a joint measurement $s:=(s_{_{1}},...,s_{_{N}}),$ and $\chi
_{_{D_{_{s}}^{(q_{s})}}}(\lambda _{1}^{(s_{_{1}})},...,\lambda
_{N}^{(s_{_{N}})})$ is an indicator function\footnote{%
See footnote 11.} of a subset $D_{_{s}}^{(q_{s})}$, we derive the following
corollary of theorem 1.

\begin{corollary}
Let an $S_{1}\times ...\times S_{N}$-setting $N$-partite correlation
experiment (\ref{1}) admit an LHV model. Then the tight linear LHV
constraint on joint probabilities:%
\begin{eqnarray}
&&\inf_{\lambda _{_{1}}\in \Lambda _{_{1}},...,\lambda _{_{N}}\in \Lambda
_{_{N}}}\text{ }\sum_{q_{s},\text{ }s}\gamma _{_{s}}^{(q_{s})}\chi
_{D_{_{s}}^{(q_{s})}}(\lambda _{1}^{(s_{_{1}})},...,\lambda
_{N}^{(s_{_{N}})})  \label{40} \\
&&  \notag \\
&\leq &\sum_{q_{s},\text{ }s}\text{ }\gamma
_{_{s}}^{(q_{s})}P_{s}(D_{s}^{(q_{s})})  \notag \\
&&  \notag \\
&\leq &\sup_{\lambda _{_{1}}\in \Lambda _{_{1}},...,\lambda _{_{N}}\in
\Lambda _{_{N}}}\text{ }\sum_{q_{s},\text{ }s}\gamma _{_{s}}^{(q_{s})}\chi
_{D_{_{s}}^{(q_{s})}}(\lambda _{1}^{(s_{_{1}})},...,\lambda
_{N}^{(s_{_{N}})}),  \notag
\end{eqnarray}%
\medskip holds for any real coefficients $\gamma _{s}^{(q_{s})}$ and any
events $D_{s}^{(q_{s})}\subseteq \Lambda _{1}^{(s_{1})}\times ...\times
\Lambda _{N}^{(s_{_{N}})},$ $q_{s}=1,...,Q_{s},$ observed under an $N$%
-partite joint measurement $s:=(s_{_{1}},...,s_{N})$ in family (\ref{1}).
\end{corollary}

\bigskip

If, for example, we take in (\ref{40}) coefficients,\ singling out only one
joint measurement: $\gamma _{\widetilde{s}}^{(q_{\widetilde{s}})}=\delta _{s,%
\widetilde{s}}$, $\forall q_{\widetilde{s}},$ and events $%
D_{s}^{(q_{s})}\subseteq $ $\Lambda _{1}^{(s_{1})}\times ...\times \Lambda
_{N}^{(s_{_{N}})}$, that are incompatible and satisfy the relation $\cup
_{q_{s}}D_{s}^{(q_{s})}=\Lambda _{1}^{(s_{1})}\times ...\times \Lambda
_{N}^{(s_{_{N}})},$ then (\ref{40}) reduces to the relation $%
\sum_{q_{_{s}}}P_{s}(D_{s}^{(q_{s})})=1,$ fulfilled under any measurement.

\section{Examples}

The general representation (\ref{8}) and its specifications\ in corollaries
1 - 4 incorporate as particular cases all Bell-type inequalities\footnote{%
On the definition of a Bell-type inequality, see the beginning of
Introduction.} for either correlation functions or joint probabilities that
have been introduced in the literature.

In this section, we explicitly demonstrate this for the most known Bell-type
inequalities. Namely, for: (1) the Clauser-Horne-Shimony-Holt (CHSH)
inequality [2] for correlation functions; (2) the Clauser-Horne (CH)
inequalities [3] for joint probabilities; (3) the Mermin-Klyshko (MK)
inequality [6 - 8] for correlation functions; (4) the Bell-type inequalities
for joint probabilities found computationally [17] by Collins and Gisin; (5)
the Bell-type inequality for joint probabilities introduced recently by
Zohren and Gill [18].

Specifying constraint (\ref{40}) for appropriate coefficients and events, it
is also easy to derive all Bell-type inequalities derived by Collins, Gisin,
Linden, Massar and Popescu in [13].

We stress that our approach allows us to derive all these inequalities in a 
\emph{new} \emph{unified manner }and also \emph{to extend} the applicability
ranges of even the well-known Bell-type inequalities.

\subsection{The Clauser-Horne-Shimony-Holt (CHSH) inequality}

For a $2\times 2$-setting bipartite correlation experiment, with real-valued
outcomes in $[-1,1]$ of any spectral type, discrete or continuous, let us
specify the tight LHV constraint (\ref{31}) with coefficients $\gamma
_{(s_{1},s_{2})}$ of the CHSH form [2]: 
\begin{equation}
(\gamma _{(s_{1},s_{2})}^{^{{\small CHSH}}})=\pm 
\begin{pmatrix}
1 & 1 \\ 
1 & -1%
\end{pmatrix}%
,  \label{41}
\end{equation}%
where minus sign may equivalently stand in any matrix cell.

Note that, in a bipartite case, two parties are traditionally named as Alice
and Bob and their measurements are usually specified by parameters $a_{i}$
and $b_{k}.$ Therefore, in case of a \emph{bipartite} correlation
experiment, we further replace our general notations of section 2 for
coefficients, outcomes and events by the following ones: 
\begin{eqnarray}
\gamma _{(s_{1},s_{2})}^{(q_{1},q_{2})} &\rightarrow &\gamma
_{_{ik}}^{^{(j,l)}}\text{, \ \ }\lambda _{1}^{(s_{1})}\rightarrow \lambda
_{1}^{(a_{i})},\text{ \ \ }\lambda _{2}^{(s_{2})}\rightarrow \lambda
_{2}^{(b_{k})},\text{ \ \ \ \ \ \ }i=1,...,S_{1},\text{ \ \ }k=1,...,S_{2},
\label{42} \\
D_{1}^{(s_{1},q_{1})} &\rightarrow &A_{i}^{(j)},\text{ \ \ }%
D_{2}^{(s_{2},q_{2})}\rightarrow B_{k}^{(l)},\text{ \ \ \ \ \ \ }%
j=1,...,Q_{1}\text{, \ \ }l=1,...,Q_{2}.  \notag
\end{eqnarray}%
Here, for concreteness, we refer site $"1"$ to Alice and site $"2"$ - to
Bob. For matrix $\gamma =(\gamma _{_{ik}}^{^{(j,l)}})\equiv (\gamma
_{ij,kl}) $ of dimension $S_{1}Q_{1}\times S_{2}Q_{2}$, the double indices $%
(i,j)$ and $(k,l)$ numerate, correspondingly, rows and columns in the order: 
\begin{eqnarray}
&&(1,1),(1,2)....,(1,Q_{1}),...,(S_{1},1),....,(S_{1},Q_{1});  \label{42'} \\
&&  \notag \\
&&(1,1),(1,2)....,(1,Q_{2}),...,(S_{2},1),....,(S_{2},Q_{2}),  \notag
\end{eqnarray}%
respectively, and element $\gamma _{_{ik}}^{^{(j,l)}}$ stands in $\gamma $
at the intersection of row $(i,j)$ and column $(k,l).$

For the CHSH coefficients (\ref{41}), the maximum of the absolute value of
the bilinear form:%
\begin{equation}
F_{2}^{^{{\small CHSH}}}(\eta _{1},\eta _{2})=\pm \left\{ \eta
_{1}^{(1)}\eta _{2}^{(1)}+\eta _{1}^{(1)}\eta _{2}^{(2)}+\eta _{1}^{(2)}\eta
_{2}^{(1)}-\eta _{1}^{(2)}\eta _{2}^{(2)}\right\}  \label{43}
\end{equation}%
over $\eta _{1}=(\eta _{1}^{(1)},\eta _{1}^{(2)})\in \{-1,1\}^{2},$ $\eta
_{2}=(\eta _{2}^{(1)},\eta _{2}^{(2)})\in \{-1,1\}^{2}$, is equal to%
\begin{equation}
\max_{(\eta _{1},\eta _{2})\in \{-1,1\}^{4}}\left\vert \text{ }F_{2}^{^{%
{\small CHSH}}}(\eta _{1},\eta _{2})\right\vert =2.  \label{44}
\end{equation}%
Substituting (\ref{44}) into (\ref{31}), we come to the following tight LHV
constraint on correlation functions:$\medskip $%
\begin{equation}
\left\vert \text{ }\langle \lambda _{1}^{(a_{1})}\lambda
_{2}^{(b_{1})}\rangle +\langle \lambda _{1}^{(a_{1})}\lambda
_{2}^{(b_{2})}\rangle +\langle \lambda _{1}^{(a_{2})}\lambda
_{2}^{(b_{1})}\rangle -\langle \lambda _{1}^{(a_{2})}\lambda
_{2}^{(b_{2})}\rangle \text{ }\right\vert _{_{LHV}}\text{ }\leq \text{ }2,
\label{45}
\end{equation}%
$\medskip $where minus sign may equivalently stand before any of four terms.
This constraint holds for outcomes in $[-1,1]$ of any spectral type,
discrete or continuous, and constitutes the Clauser-Horne-Shimony-Holt
(CHSH) inequality, derived originally in [2] for two $\pm 1$-valued outcomes
per site and further proved [5] by Bell to hold for any outcomes $|\lambda
_{1}^{(a_{i})}|,$ $|\lambda _{2}^{(b_{k})}|$ $\leq 1,$ $i,k=1,2.$

\subsection{The Clauser-Horne (CH) inequalities}

For a $2\times 2$-setting bipartite correlation experiment, let us specify
constraint (\ref{9}) with the CHSH coefficients (\ref{41}) and $\pm 1$%
-valued functions 
\begin{eqnarray}
\phi _{1}^{(i)}(\lambda _{1}^{(a_{i})}) &=&2\chi _{A_{i}}(\lambda
_{1}^{(a_{i})})-1,\text{ \ \ \ \ }i=1,2,  \label{46} \\
\phi _{2}^{(k)}(\lambda _{2}^{(b_{k})}) &=&2\chi _{B_{k}}(\lambda
_{2}^{(b_{k})})-1,\text{ \ \ \ \ }k=1,2,  \notag
\end{eqnarray}%
where $A_{i}\subseteq \Lambda _{1}^{(a_{i})}$ and $B_{k}\subseteq \Lambda
_{2}^{(s_{2})}$ are any events observed by Alice and Bob under the
corresponding measurements.

For these functions, the product expectations take the form:%
\begin{eqnarray}
\left\langle \phi _{1}^{(i)}(\lambda _{1}^{(a_{i})})\phi _{2}^{(k)}(\lambda
_{2}^{(b_{k})})\right\rangle &=&1+4P_{(a_{i},b_{k})}(A_{i}\times
B_{k})-2P_{(a_{i},b_{k})}(A_{i}\times \Lambda _{2}^{(b_{k})})  \label{47} \\
&&-2P_{(a_{i},b_{k})}(\Lambda _{1}^{(a_{i})}\times B_{k}),  \notag
\end{eqnarray}%
and ranges (\ref{11}) satisfy the relation: $\Phi _{1},$ $\Phi _{2}\subseteq
\{-1,1\}^{2}$. The latter implies: 
\begin{eqnarray}
\max_{\xi _{1}\in \Phi _{1},\text{ }\xi _{2}\in \Phi _{2}}F_{2}^{^{{\small %
CHSH}}}(\xi _{1},\xi _{2}) &\leq &\max_{(\xi _{1},\xi _{2})\text{ }\in \text{
}\{-1,1\}^{4}}F_{2}^{^{{\small CHSH}}}(\xi _{1},\xi _{2}),  \label{48} \\
\min_{\xi _{1}\in \Phi _{1},\text{ }\xi _{2}\in \Phi _{2}}F_{2}^{{\small CHSH%
}}(\xi _{1},\xi _{2}) &\geq &\min_{(\xi _{1},\xi _{2})\text{ }\in \text{ }%
\{-1,1\}^{4}}F_{2}^{^{{\small CHSH}}}(\xi _{1},\xi _{2}).  \notag
\end{eqnarray}%
Taking into account (\ref{30}), (\ref{44}), we have:%
\begin{eqnarray}
\max_{(\xi _{1},\xi _{2})\text{ }\in \text{ }\{-1,1\}^{4}}F_{2}^{^{{\small %
CHSH}}}(\xi _{1},\xi _{2}) &=&-\min_{(\xi _{1},\xi _{2})\text{ }\in \text{ }%
\{-1,1\}^{4}}F_{2}^{^{{\small CHSH}}}(\xi _{1},\xi _{2})  \label{49} \\
&&  \notag \\
&=&\max_{(\xi _{1},\xi _{2})\text{ }\in \text{ }\{-1,1\}^{4}}\left\vert
F_{2}^{^{{\small CHSH}}}(\xi _{1},\xi _{2})\right\vert  \notag \\
&&  \notag \\
&=&2.  \notag
\end{eqnarray}%
Substituting (\ref{47}) - (\ref{49}) into (\ref{9}) and noting that, for 
\emph{EPR local} measurements of Alice and Bob, the marginal probabilities
in (\ref{47}) have the form\footnote{%
See condition (\ref{3}) and notation (\ref{4}).}:%
\begin{equation}
P_{(a_{i},b_{k})}(A_{i}\times \Lambda _{2}^{(b_{k})})=P_{1}^{(a_{i})}(A_{i}),%
\text{ \ \ }P_{(a_{i},b_{k})}(\Lambda _{1}^{(a_{i})}\times
B_{k})=P_{2}^{(b_{k})}(B_{k}),  \label{50}
\end{equation}%
we come to the following LHV constraint on joint probabilities:\medskip 
\begin{eqnarray}
-1 &\leq &P_{(a_{1},b_{1})}(A_{1}\times B_{1})+P_{(a_{1},b_{2})}(A_{1}\times
B_{2})+P_{(a_{2},b_{1})}(A_{2}\times B_{1})  \label{51} \\
&&-P_{(a_{2},b_{2})}(A_{2}\times
B_{2})-P_{1}^{(a_{1})}(A_{1})-P_{2}^{(b_{1})}(B_{1})\leq 0.  \notag
\end{eqnarray}%
This LHV constraint is valid for\emph{\ any} events $A_{i}\subseteq \Lambda
_{1}^{(a_{i})},$ $B_{k}\subseteq \Lambda _{2}^{(b_{k})},$ observed by Alice
and Bob under measurements $a_{i}$, $i=1,2,$ and $b_{k},$ $k=1,2,$
respectively, and corresponds to the Clauser-Horne (CH) inequalities [3] on
joint probabilities.

We stress that, in (\ref{51}), outcome events may be \emph{arbitrary}, in
particular, certain: $A_{i}=\Lambda _{1}^{(a_{i})}$, $B_{k}=\Lambda
_{2}^{(b_{k})},$ or impossible: $A_{i}=\varnothing ,$ $B_{k}=\varnothing .$
This implies that, in the form (\ref{51}), the CH inequalities incorporate 
\emph{as particular cases }all positive probability relations considered in
the literature\footnote{%
See, for example, in [4].} usually separately. If, for example, $%
A_{2}=B_{1}=\varnothing $ then (\ref{51}) reduces to the positive
probability relation $-1\leq $ $P_{(a_{1},b_{2})}(A_{1}\times B_{2})$ $%
-P_{1}^{(a_{1})}(A_{1})\leq 0$, fulfilled under any bipartite joint
measurement.

Note also that the CH inequalities (\ref{51}) are equivalent\footnote{%
In the sense that the validity of the CHSH inequality on correlation
functions implies the validity of the CH inequalities on joint probabilities
and vice versa.} to the CHSH inequality (\ref{45}) only in case of two $\pm
1 $-valued outcomes at each site and the choice in (\ref{51}) of uncertain
possible events, say $A_{i}=\{1\},$ $B_{k}=\{1\},$ for any $i,k\in \{1,2\}.$

\subsection{The Mermin-Klyshko\ (MK) inequality}

For a $\underbrace{2\times ...\times 2}_{N}$-setting $N$-partite correlation
experiment, with outcomes in $[-1,1]$ of any spectral type, discrete or
continuous, let us specify constraint (\ref{31}) with coefficients $\gamma
_{_{(s_{1},...,s_{N})}}$ defined by recursion: 
\begin{equation}
\gamma _{(s_{1},...,s_{n-1},s_{n})}=\gamma _{(s_{1},...,s_{n-1})}+(\delta
_{s_{n,}1}-\delta _{s_{n},2})\gamma _{(\overline{s_{1}},...,\overline{s_{n-1}%
})},\text{ \ \ }3\leq n\leq N,  \label{52}
\end{equation}%
where $(\gamma _{(s_{_{1}},s_{_{2}})})=(\gamma _{(s_{_{1}},s_{_{2}})}^{^{%
{\small CHSH}}})=%
\begin{pmatrix}
1 & 1 \\ 
1 & -1%
\end{pmatrix}%
$ and $\overline{s_{n}}$ is the element of set $\{1,2\}\backslash \{s_{n}\}.$

In order to find the maximum of the absolute value of the $N$-linear form%
\begin{equation}
F_{N}^{(\gamma )}(\eta _{1},...,\eta
_{N})=\tsum\limits_{s_{1},...,s_{_{N}}=1,2}\gamma _{(s_{1},...,s_{_{N}})}%
\text{ }\eta _{1}^{(s_{1})}\cdot ...\cdot \eta _{N}^{(s_{_{N}})}  \label{53}
\end{equation}%
over vectors $\eta _{1}\in \{-1,1\}^{2},$ $...,$ $\eta _{n}\in \{-1,1\}^{2},$
let us introduce $n$-linear forms, corresponding to $n$-th step in recursion
(\ref{52}):\ 
\begin{eqnarray}
F_{n}^{^{(\gamma )}}(\eta _{1},...,\eta _{n})
&:&=\sum_{s_{1},...,s_{n}=1,2}\gamma _{(s_{1},...,s_{n})}\text{ }\eta
_{1}^{(s_{1})}\cdot ...\cdot \eta _{n}^{(s_{n})},  \label{54} \\
\overline{F}_{n}^{^{(\gamma )}}(\eta _{1},...,\eta _{n})
&:&=\sum_{s_{1},...,s_{n}=1,2}\gamma _{(\overline{s_{1},}...,\overline{s_{n}}%
)}\text{ }\eta _{1}^{(s_{1})}\cdot ...\cdot \eta _{n}^{(s_{n})}.  \notag
\end{eqnarray}%
Substituting (\ref{52}) into (\ref{54}), we have:%
\begin{eqnarray}
F_{n}^{(\gamma )}(\eta _{1},...,\eta _{n}) &=&(\eta _{n}^{(1)}+\eta
_{n}^{(2)})\text{ }F_{n-1}^{(\gamma )}(\eta _{1},...,\eta _{n-1})  \label{55}
\\
&&+(\eta _{n}^{(1)}-\eta _{n}^{(2)})\text{ }\overline{F}_{n-1}^{(\gamma
)}(\eta _{1},...,\eta _{n-1}),\text{ \ \ }n\geq 3,  \notag
\end{eqnarray}%
where 
\begin{eqnarray}
F_{2}^{(\gamma )}(\eta _{1},\eta _{2}) &=&\eta _{1}^{(1)}\eta
_{2}^{(1)}+\eta _{1}^{(1)}\eta _{2}^{(2)}+\eta _{1}^{(2)}\eta
_{2}^{(1)}-\eta _{1}^{(2)}\eta _{2}^{(2)},  \label{56} \\
&&  \notag \\
\overline{F}_{2}^{(\gamma )}(\eta _{1},\eta _{2}) &=&-\eta _{1}^{(1)}\eta
_{2}^{(1)}+\eta _{1}^{(1)}\eta _{2}^{(2)}+\eta _{1}^{(2)}\eta
_{2}^{(1)}+\eta _{1}^{(2)}\eta _{2}^{(2)}.  \notag
\end{eqnarray}%
Taking into account (\ref{44}), (\ref{55}), we prove by induction in $n$ the
following relation:\medskip 
\begin{equation}
\max_{(\eta _{1},...,\eta _{N})\in \{-1,1\}^{2N}}\text{ }\left\vert
F_{N}^{(\gamma )}(\eta _{1},...,\eta _{N})\right\vert =2^{N-1},\text{ \ \ \
\ }N\geq 2.  \label{57}
\end{equation}%
\medskip

Substituting (\ref{57}) into (\ref{31}), we come to the following $%
\underbrace{2\times ...\times 2}_{N}$-setting tight LHV\ constraint on the
full correlation functions:\medskip\ 
\begin{equation}
\left\vert \sum_{s_{1},...,s_{_{N}}\in \{1,2\}}\gamma
_{(s_{1},...,s_{_{N}})}\left\langle \lambda _{1}^{(s_{1})}\cdot ...\cdot
\lambda _{N}^{(s_{_{N}})}\right\rangle _{_{LHV}}\right\vert \leq 2^{N-1},
\label{58}
\end{equation}%
\medskip where coefficients $\gamma _{(s_{1},...,s_{_{N}})}$ are given by (%
\ref{52}). For $N=2$, this inequality reduces to the CHSH inequality (\ref%
{45}).

Let us now specify constraint (\ref{58}) for a $\underbrace{2\times
...\times 2}_{N}$-setting correlation experiment, with outcomes in $[-1,1]$
of any spectral type, discrete or continuous, performed on a quantum state $%
\rho $ on a complex separable Hilbert space $\mathcal{H}_{1}\otimes
...\otimes \mathcal{H}_{N},$ possibly infinite dimensional.

In the quantum case\footnote{%
Here, $\mathrm{M}_{n}^{(s_{n})}(d\lambda _{n}^{(s_{n})})$ is a positive
operator-valued (POV) measure describing $s_{n}$-th measurement at $n$-th
site, see, for example, section 3.1 in [20].}, 
\begin{eqnarray}
\langle \lambda _{1}^{(s_{1})}\cdot ...\cdot \lambda
_{N}^{(s_{_{N}})}\rangle _{\rho } &=&\dint \lambda _{1}^{(s_{1})}\cdot
...\cdot \lambda _{N}^{(s_{_{N}})}\mathrm{tr}[\rho \{\mathrm{M}%
_{1}^{(s_{1})}(\mathrm{d}\lambda _{1}^{(s_{1})})\otimes ...\otimes \mathrm{M}%
_{N}^{(s_{N})}(\mathrm{d}\lambda _{N}^{(s_{_{N}})})\}]  \notag \\
&=&\mathrm{tr}[\rho (X_{1}^{(s_{1})}\otimes ...\otimes X_{N}^{(s_{_{N}})})],
\label{59}
\end{eqnarray}%
where 
\begin{equation}
X_{n}^{(s_{n})}=\int \lambda _{n}^{(s_{n})}\mathrm{M}_{n}^{(s_{n})}(\mathrm{d%
}\lambda _{n}^{(s_{n})})  \label{60}
\end{equation}%
is a bounded quantum observable on $\mathcal{H}_{n}$, observed under $s_{n}$%
-th measurement at $n$-th site and with operator norm $||X_{n}^{(s_{n})}||$ $%
\leq 1$. If a Hilbert space $\mathcal{H}_{n}$, corresponding to $n$-th site,
is infinite dimensional then observables $X_{n}^{(s_{n})},$ $s_{n}=1,2,$%
\emph{\ }may be of any spectral type, discrete or continuous.

From (\ref{59}), (\ref{52}) it follows that, in the quantum case,%
\begin{equation}
\sum_{s_{1},...,s_{_{N}}}\gamma _{(s_{1},...,s_{_{N}})}\langle \text{ }%
\lambda _{1}^{(s_{1})}\cdot ...\cdot \lambda _{N}^{(s_{N})}\text{ }\rangle
_{\rho }=\mathrm{tr}[\rho \mathcal{B}_{N}],  \label{61}
\end{equation}%
where $\mathcal{B}_{N}$ is the bounded quantum observable\footnote{$\mathcal{%
B}_{N}$ represents a generalization of the so-called Bell operator for spin
measurements on $N$ qubits.} on $\mathcal{H}_{1}\otimes ...\otimes \mathcal{H%
}_{N}$, defined by recursion$\medskip $ 
\begin{eqnarray}
\mathcal{B}_{n} &=&(X_{n}^{(1)}+X_{n}^{(2)})\otimes \mathcal{B}%
_{n-1}+(X_{n}^{(1)}-X_{n}^{(2)})\otimes \widetilde{\mathcal{B}}_{n-1},\text{
\ \ \ \ }2\leq n\leq N,  \label{62} \\
\mathcal{B}_{1} &=&X_{1}^{(1)},\text{ \ \ }\widetilde{\mathcal{B}}%
_{1}=X_{1}^{(2)},  \notag
\end{eqnarray}%
where $\widetilde{\mathcal{B}}_{n}$ results from $\mathcal{B}_{n}$ by
interchanging all $X_{k}^{(s_{k})}$ to $X_{k}^{(\overline{s_{k}})},$ $%
s_{k}=1,2;$ $k=1,...n.$

Substituting (\ref{61}) into (\ref{58}), we come to the quantum version%
\begin{equation}
\left\vert \text{ }\mathrm{tr}[\rho \mathcal{B}_{N}]\text{ }\right\vert
_{_{LHV}}\leq 2^{N-1}  \label{63}
\end{equation}%
of the tight LHV constraint (\ref{58}). By its form, this quantum LHV
constraint coincides with the Mermin-Klyshko (MK) inequality, derived
originally\footnote{%
Mermin's inequality [6] and the similar inequality of Ardehali [7]
distinguish between even and odd values of $N$. For an odd $N,$ the
magnitude of the maximal violation of Mermin's inequality in a quantum case
is higher than that of Ardehali. For an even $N,$ the situation is opposite.
Belinskii and Klyshko [8] proposed the single inequality, which is maximally
violated, in comparison with those in [6, 7], for any $N$, even or odd. This
inequality is usually referred to as the Mermin-Klyshko inequality.} [6-8]
for the LHV description of spin measurements on $N$ qubits and still
discussed in the literature (see, for example, in [10]) only for a $N$%
-partite case with two dichotomic observables per site.

Our derivation of (\ref{63}) shows that, for an $N$-partite quantum state $%
\rho $ on $\mathcal{H}_{1}\otimes ...\otimes \mathcal{H}_{N}$, possibly
infinite dimensional, admitting the $\underbrace{2\times ...\times 2}_{N}$%
-setting LHV description\footnote{%
See section 5 of [20].}, the MK inequality holds for \emph{arbitrary }two
quantum observables per site, not necessarily dichotomic. If $\mathcal{H}%
_{n} $ is infinite dimensional then quantum observables measured at $n$-th
site may be of any spectral type, discrete or continuous.

\subsection{The Collins-Gisin inequalities}

Let us now demonstrate that the tight LHV constraint (\ref{38}) on joint
probabilities incorporate as particular cases the \emph{extreme} bipartite
Bell-type inequalities found by Collins and Gisin [17] computationally. For
short, we consider here the derivation of only two inequalities reported in
[17].

For a $4\times 4$-setting bipartite correlation experiment, with at least 
\emph{two} outcomes per site, let us specify (\ref{38}) with $Q_{1}=Q_{2}=1,$
matrix%
\begin{equation}
\gamma =(\gamma _{ik})=%
\begin{pmatrix}
1 & 1 & 1 & 1 \\ 
1 & 1 & 1 & -1 \\ 
1 & 1 & -1 & 0 \\ 
1 & -1 & 0 & 0%
\end{pmatrix}
\label{64}
\end{equation}%
and vectors 
\begin{equation}
\gamma _{1}=(-1,0,0,0),\text{ \ \ \ \ }\gamma _{2}=(-3,-2,-1,0).  \label{65}
\end{equation}%
In this case, sets (\ref{37}) take the form: $\Xi _{1}=\Xi _{2}=\{0,1\}^{4}$%
, and maximum%
\begin{equation}
\max_{\eta _{1}\in \{0,1\}^{4},\text{ }\eta _{2}\in \{0,1\}^{4}}\left\{ 
\text{ }(\eta _{1},\gamma \eta _{2})+(\eta _{1},\gamma _{1})+(\eta
_{2},\gamma _{2})\right\} =0,  \label{66}
\end{equation}%
achieved at, for example, $\eta _{1}=(1,1,1,1),$ $\eta _{2}=(1,1,1,1).$%
Substituting (\ref{64}) - (\ref{66}) into the right-hand side inequality of (%
\ref{38}), we come to the tight LHV constraint:$\medskip $ 
\begin{eqnarray}
&&\sum_{i,k}\gamma _{ik}P_{(a_{i},b_{k})}(A_{i}\times B_{k})\text{ }%
-P_{1}^{(a_{1})}(A_{1})  \label{67} \\
&&-3P_{2}^{(b_{1})}(B_{1})-2P_{2}^{(b_{2})}(B_{2})-P_{2}^{(b_{3})}(B_{3})%
\leq 0,  \notag
\end{eqnarray}%
$\medskip $corresponding to the extreme Bell-type inequality $I_{4422}\leq 0$%
, introduced in [17, Eq. (38)], and valid for any events: $A_{i}\subset
\Lambda _{1}^{(a_{i})}$, $A_{i}\neq \varnothing ,$ $B_{k}\subset \Lambda
_{2}^{(b_{k})}$, $B_{k}\neq \varnothing ,$ observed by Alice and Bob under
the corresponding measurements.

For a $2\times 2$-setting bipartite correlation experiment, with at least 
\emph{three} outcomes per site, let us also specify (\ref{38}) with $%
Q_{1}=Q_{2}=2,$ vectors 
\begin{equation}
\gamma _{1}=\gamma _{2}=(-1,-1,0,0)  \label{72}
\end{equation}%
and matrix%
\begin{equation}
\gamma =(\gamma _{ik}^{(j,l)})=%
\begin{pmatrix}
1 & 1 & 0 & 1 \\ 
1 & 0 & 1 & 1 \\ 
0 & 1 & 0 & -1 \\ 
1 & 1 & -1 & -1%
\end{pmatrix}%
,  \label{74}
\end{equation}%
where element $\gamma _{ik}^{(j,l)}$ stands\footnote{%
See also (\ref{42'}).} in $\gamma $ at the intersection of row $(i,j)$ and
column $(k,l)$.

In this case, sets (\ref{37}) are given by:%
\begin{eqnarray}
\Xi _{1} &=&\{\eta _{1}\in \{0,1\}^{4}\mid \tsum\limits_{j=1,2}\eta
_{1}^{(i,j)}\in \{0,1\},\text{ \ \ }i=1,2\},  \label{75} \\
\Xi _{2} &=&\{\eta _{2}\in \{0,1\}^{4}\mid \tsum\limits_{l=1,2}\eta
_{2}^{(k,l)}\in \{0,1\},\text{ \ \ }k=1,2\},  \notag
\end{eqnarray}%
and%
\begin{equation}
\max_{\eta _{1}\in \Xi _{1},\text{ }\eta _{2}\in \Xi _{2}}\text{ }\left\{ 
\text{ }(\eta _{1},\gamma \eta _{2})+(\eta _{1},\gamma _{1})+(\eta
_{2},\gamma _{2})\right\} =0,  \label{76}
\end{equation}%
\smallskip achieved at, for example, $\eta _{1}=(1,0,0,1)$ and $\eta
_{2}=(1,0,0,0).$

Substituting (\ref{74}) - (\ref{76}) into the right-hand side inequality of (%
\ref{38}), we derive the tight LHV constraint 
\begin{eqnarray}
&&\sum_{i,j,k,l}\gamma _{ik}^{(j,l)}P_{(a_{i},b_{k})}(A_{i}^{(j)}\times
B_{k}^{(l)})  \label{77} \\
&&-P_{1}^{(a_{1})}(A_{1}^{(1)})-P_{1}^{(a_{1})}(A_{1}^{(2)})-P_{2}^{(b_{1})}(B_{1}^{(1)})-P_{2}^{(b_{1})}(B_{1}^{(2)})%
\text{ }\leq \text{ }0,  \notag
\end{eqnarray}%
\smallskip corresponding to the extreme Bell-type inequality $I_{2233}\leq
0, $ introduced analytically in [13, 14] and further confirmed
computationally in [17, Eq. (39)]. This inequality is valid for any two
incompatible events%
\begin{equation}
A_{i}^{(j)}\subset \Lambda _{i}^{(a_{i})},\text{ \ \ }A_{i}^{(j)}\neq
\varnothing ,\text{ \ \ }j=1,2,\text{ \ \ }A_{i}^{(1)}\cap
A_{i}^{(2)}=\varnothing ,\text{ \ \ }A_{i}^{(1)}\cup A_{i}^{(2)}\neq \Lambda
_{i}^{(a_{i})},  \label{78}
\end{equation}%
observed by Alice under measurement $a_{i},$ $i=1,2,$ and any two
incompatible events%
\begin{equation}
B_{k}^{(l)}\subseteq \Lambda _{2}^{(b_{k})},\text{ \ \ }B_{k}^{(l)}\neq
\varnothing ,\ \ \ l=1,2,\text{ \ \ \ }B_{k}^{(1)}\cap
B_{k}^{(2)}=\varnothing ,\text{ \ \ }B_{k}^{(1)}\cup B_{k}^{(2)}\neq \Lambda
_{2}^{(b_{k})},  \label{79}
\end{equation}%
observed by Bob under measurement $b_{k},$ $k=1,2.$

\subsection{The Zohren-Gill inequality}

Finally, consider a $2\times 2$-setting bipartite correlation experiment
with $K$ real-valued outcomes per site: $\lambda _{1}^{(s_{1})},$ $\lambda
_{2}^{(s_{2})}\in \Lambda =\{1,...,K\},$ where $2\leq K\leq \infty .$

For this case, let us specify the tight linear LHV constraint (\ref{40})
with $\gamma _{(s_{1},s_{2})}=1,$ $q_{(s_{1},s_{2})}=1,$ $\forall
s_{1},s_{2}\in \{1,2\},$ and events: 
\begin{eqnarray}
D_{(s_{1},s_{2})} &=&\{\lambda _{2}^{(s_{2})}>\lambda
_{1}^{(s_{1})}\}\subset \Lambda \times \Lambda ,\text{ \ \ if \ }%
s_{1}=s_{2}\in \{1,2\},  \label{80} \\
D_{(s_{1},s_{2})} &=&\{\lambda _{1}^{(s_{1})}>\lambda
_{2}^{(s_{1})}\}\subset \Lambda \times \Lambda ,\text{ \ \ if}\ \ s_{1}\neq
s_{2}\in \{1,2\}.  \notag
\end{eqnarray}%
We have: 
\begin{eqnarray}
\chi _{_{_{D_{(s_{1},s_{2})}}}}(\lambda _{1}^{(s_{_{1}})},\lambda
_{2}^{(s_{_{2}})}) &=&\theta (\lambda _{2}^{(s_{2})}-\lambda _{1}^{(s_{1})}),%
\text{ \ \ if \ }s_{1}=s_{2}\in \{1,2\},  \label{82} \\
\chi _{_{_{D_{(s_{1},s_{2})}}}}(\lambda _{1}^{(s_{_{1}})},\lambda
_{2}^{(s_{_{2}})}) &=&\theta (\lambda _{1}^{(s_{1})}-\lambda _{2}^{(s_{2})}),%
\text{ \ \ if \ }s_{1}\neq s_{2}\in \{1,2\},  \notag
\end{eqnarray}%
where $\theta (x-y)=1,$ for $x>y,$ and $\theta (x-y)=0,$ for $x\leq y.$
Substituting (\ref{82}) into the left-hand side inequality of the LHV
constraint (\ref{40}), we derive the following expression for 
\begin{equation}
\inf \left\{ \theta (\lambda _{2}^{(1)}-\lambda _{1}^{(1)})+\theta (\lambda
_{1}^{(1)}-\lambda _{2}^{(2)})+\theta (\lambda _{2}^{(2)}-\lambda
_{1}^{(2)})+\theta (\lambda _{1}^{(2)}-\lambda _{2}^{(1)})\right\} =1
\label{83}
\end{equation}%
over all $\lambda _{1}^{(1)},\lambda _{1}^{(2)},\lambda _{2}^{(1)},\lambda
_{2}^{(2)}\in \{1,...,K\},$ and, therefore, the following tight LHV
constraint%
\begin{eqnarray}
P_{(1,1)}(\{\lambda _{2}^{(1)} &>&\lambda _{1}^{(1)}\})\text{ \ }+\text{ \ }%
P_{(1,2)}(\{\lambda _{1}^{(1)}>\lambda _{2}^{(2)}\})  \label{84'} \\
+\text{ \ }P_{(2,2)}(\{\lambda _{2}^{(2)} &>&\lambda _{1}^{(2)}\})\text{ \ }+%
\text{ \ }P_{(2,1)}(\{\lambda _{1}^{(2)}>\lambda _{2}^{(1)}\})\text{ \ \ }%
\geq \text{ }1,  \notag
\end{eqnarray}%
which is valid for any number $K$ of outcomes per site, in particular, for
inifinitely many outcomes $(K=\infty )$ at each site. This tight LHV
constraint constitutes the Bell-type inequality derived quite differently in
[18].

\section{Conclusions}

In the present paper, which is a sequel to [20], we have introduced in
rigorous mathematical terms a single general representation for all tight
linear LHV constraints arising under an $S_{1}\times ...\times S_{N}$%
-setting $N$-partite correlation experiment with outcomes of any spectral
type, discrete or continuous. For correlation functions and joint
probabilities, this representation is formulated in terms of multilinear
forms and this allows us:

\begin{enumerate}
\item[$\bullet $] to prove in a general setting that the form of any
correlation Bell-type inequality does not depend on a spectral type of
outcomes at different sites, in particular, on their numbers and is
determined only by extremal values of outcomes at each site,

\item[$\bullet $] to specify the general form of bounds in Bell-type
inequalities for joint probabilities;

\item[$\bullet $] to present the new concise proofs for all the most known
Bell-type inequalities introduced in the literature ever since the seminal
publication of Bell [1] and also to extend the applicability ranges of some
of these inequalities.
\end{enumerate}

Note that the LHV constraints, reproduced in sections 3.1-3.4, are not only
tight, but, as it is proved in [4, 10, 17], respectively, each of these
inequalities is extreme for the corresponding setting of a correlation
experiment. However, for an arbitrary multipartite case, there does not
still exist an effective general way to single out extreme Bell-type
inequalities. Though the polytope approach is very useful from the
descriptive-geometrical point of view, there is no much sense in finding of
extreme Bell-type inequalities by listing of a huge number of faces of a
highly dimensional polytope whereas many of these faces correspond to
trivial probabilistic constraints while others can be subdivided into only a
few classes different by their form.

The approach, introduced in the present paper, is based on general
properties of multilinear forms and this points to a possibility of a new
direction in finding of extreme Bell-type inequalities for an arbitrary
multipartite case. This problem will be analysed in our further publications.

\section{Appendix}

\begin{proof}[Proof of Lemma 1]
For the real-valued function%
\begin{equation}
W(\eta ):=\sum_{\substack{ 1\leq n_{1}<...<n_{M}\leq N,\text{ }  \\ %
M=1,...,N }}F_{M}^{^{(\gamma )}}(\eta _{n_{1}},...,\eta _{n_{M}})  \tag{A1}
\end{equation}%
continuous on $\mathbb{R}^{S_{1}+...+S_{N}}$, its supremum and infimum over $%
\eta =(\eta _{1},...,\eta _{N})\in \Lambda _{1}\times ...\times \Lambda
_{N}\subseteq \lbrack -1,1]^{S_{1}+...+S_{N}}$ have the form:%
\begin{eqnarray}
\sup_{\eta \in \Lambda _{1}\times ...\times \Lambda _{N}}W(\eta )
&=&\sup_{\eta \in \overline{\Lambda }_{1}\times ...\times \overline{\Lambda }%
_{_{N}}}W(\eta ),  \TCItag{A2} \\
\inf_{\eta \in \Lambda _{1}\times ...\times \Lambda _{N}}W(\eta )
&=&\inf_{\eta \in \overline{\Lambda }_{1}\times ...\times \overline{\Lambda }%
_{_{N}}}W(\eta ),  \notag
\end{eqnarray}%
where%
\begin{eqnarray}
\sup_{\eta \in \overline{\Lambda }_{1}\times ...\times \overline{\Lambda }%
_{_{N}}}W(\eta ) &=&\max_{\eta \in \overline{\Lambda }_{1}\times ...\times 
\overline{\Lambda }_{_{N}}}W(\eta ),  \TCItag{A3} \\
\inf_{\eta \in \overline{\Lambda }_{1}\times ...\times \overline{\Lambda }%
_{_{N}}}W(\eta ) &=&\min_{\eta \in \overline{\Lambda }_{1}\times ...\times 
\overline{\Lambda }_{_{N}}}W(\eta ).  \notag
\end{eqnarray}%
Therefore,%
\begin{eqnarray}
\sup_{\eta \in \Lambda _{1}\times ...\times \Lambda _{N}}W(\eta )
&=&\max_{\eta \in \overline{\Lambda }_{1}\times ...\times \overline{\Lambda }%
_{_{N}}}W(\eta ),  \TCItag{A4} \\
\inf_{\eta \in \Lambda _{1}\times ...\times \Lambda _{N}}W(\eta )
&=&\min_{\eta \in \overline{\Lambda }_{1}\times ...\times \overline{\Lambda }%
_{_{N}}}W(\eta ).  \notag
\end{eqnarray}%
From relation (\ref{25'}) it follows: 
\begin{eqnarray}
\max_{\eta \in \{-1,1\}^{d}}W(\eta ) &\leq &\max_{\eta \in \overline{\Lambda 
}_{1}\times ...\times \overline{\Lambda }_{_{N}}}W(\eta )\leq \max_{\eta \in
\lbrack -1,1]^{d}}W(\eta ),  \TCItag{A5} \\
&&  \notag \\
\min_{\eta \in \lbrack -1,1]^{d}}W(\eta ) &\leq &\min_{\eta \in \overline{%
\Lambda }_{1}\times ...\times \overline{\Lambda }_{_{N}}}W(\eta )\leq
\min_{\eta \in \{-1,1\}^{d}}W(\eta ),  \notag
\end{eqnarray}%
where $d=S_{1}+...+S_{N}$. Note that $\eta =(\eta _{1}^{(1)},...,\eta
_{1}^{(S_{1})},...,\eta _{N}^{(1)},...,\eta _{N}^{(S_{_{N}})})\in \mathbb{R}%
^{d}$ and function $W(\eta )$ is twice continuously differentiable on $%
\mathbb{R}^{d}$ with the second partial derivatives%
\begin{equation}
\frac{\partial ^{2}W(\eta )}{\partial (\eta _{n}^{(s_{n})})^{2}}=0.  \tag{A6}
\end{equation}%
Therefore, function $W(\eta ),$ $\eta \in \mathbb{R}^{d}$, is harmonic%
\footnote{%
On this notion, see any textbook on equations of mathematical physics.}.
From the maximum principle for harmonic functions it follows that the
maximum and the minimum of function $W(\eta )$ in hypercube $\mathbb{V}%
_{d}:=[-1,1]^{d}\subset \mathbb{R}^{d}$ are reached on boundary $\Gamma _{d}$
of $\mathbb{V}_{d}$, that is:%
\begin{equation}
\max_{\eta \in \lbrack -1,1]^{d}}W(\eta )=\max_{\eta \in \Gamma _{d}}W(\eta
),\text{ \ \ \ \ \ \ \ }\min_{\eta \in \lbrack -1,1]^{d}}W(\eta )=\min_{\eta
\in \Gamma _{d}}W(\eta ).  \tag{A7}
\end{equation}%
Since boundary $\Gamma _{d}\ $of $\mathbb{V}_{d}$ represents the union of $%
(d-1)$-dimensional hypercubes $\mathbb{V}_{d-1}^{(k)}$, $k=1,...,2d,$ the
right-hand sides of relations (A7) are given by:%
\begin{eqnarray}
\max_{\eta \in \Gamma _{d}}W(\eta ) &=&\max_{k=1,...,2d}\text{ }\{\max_{\eta
\in \mathbb{V}_{d-1}^{(k)}}W(\eta )\},  \TCItag{A8} \\
\text{\ }\min_{\eta \in \Gamma _{d}}W(\eta ) &=&\min_{k=1,...,2d}\text{ }%
\{\min_{\eta \in \mathbb{V}_{d-1}^{(k)}}W(\eta )\}.  \notag
\end{eqnarray}

Further, on each $(d-1)$-dimensional hypercube $\mathbb{V}_{d-1}^{(k)},$
function $W(\eta )$ $|_{\mathbb{V}_{d-1}^{(k)}},$ depending on $(d-1)$
components of $\eta ,$ is harmonic and, therefore, reaches its maximum
(minimum) on boundary $\Gamma _{d-1}^{(k)}$ of $\mathbb{V}_{d-1}^{(k)}.$ The
latter, in turn, consists, of $(d-2)$-dimensional hypercubes $\mathbb{V}%
_{d-2}^{(m)}$. Since, in total, boundary $\Gamma _{d}$ contains $4d(d-1)$ of 
$(d-2)$-dimensional hypercubes $\mathbb{V}_{d-2}^{(m)}$, relation (A8)
reduces to:%
\begin{eqnarray}
\max_{\eta \in \Gamma _{d}}W(\eta ) &=&\max_{k=1,...,2d}\text{ }\{\max_{\eta
\in \mathbb{V}_{d-1}^{(k)}}W(\eta )\}  \TCItag{A9} \\
&=&\max_{m=1,...,4d(d-1)}\text{ }\{\max_{\eta \in \mathbb{V}%
_{d-2}^{(m)}}W(\eta )\},  \notag
\end{eqnarray}%
with a similar relation for minimum.

Recall that the number of $l$-dimensional hypercubes on the boundary $\Gamma
_{d}$ is equal to 
\begin{equation}
\frac{d!}{(d-l)!}2^{d-l},  \tag{A10}
\end{equation}%
in particular, $d\cdot 2^{d-1}$ edges ($"1"$- dimensional hypercubes) and $%
2^{d}$ vertices ($"0"$- dimensional hypercubes).

Continuing to reduce the dimension of hypercubes in formula (A9), we finally
come to the maximum (minimum) over all $"0"$-dimensional hypercubes, that
is, over set $\{-1,1\}^{d}$ of all $2^{d}$ vertices of hypercube $\mathbb{V}%
_{d}=[-1,1]^{d}$. Thus:%
\begin{eqnarray}
\max_{\eta \in \lbrack -1,1]^{d}}W(\eta ) &=&\max_{\eta \in
\{-1,1\}^{d}}W(\eta ),  \TCItag{A11} \\
\min_{\eta \in \lbrack -1,1]^{d}}W(\eta ) &=&\min_{\eta \in
\{-1,1\}^{d}}W(\eta ).  \notag
\end{eqnarray}%
From (A4), (A5) and (A11) it follows:%
\begin{eqnarray}
\sup_{\eta \in \Lambda _{1}\times ...\times \Lambda _{N}}W(\eta )
&=&\max_{\eta \in \{-1,1\}^{d}}W(\eta ),  \TCItag{A12} \\
\text{\ }\inf_{\eta \in \Lambda _{1}\times ...\times \Lambda _{N}}W(\eta )
&=&\min_{\eta \in \{-1,1\}^{d}}W(\eta ).  \notag
\end{eqnarray}%
This proves the statement of lemma 1.
\end{proof}

\end{document}